\documentclass[pre,10pt,twocolumn,superscriptaddress,showpacs]{revtex4-2}
\usepackage[T1]{fontenc}
\usepackage{amsmath}
\usepackage{latexsym}
\usepackage{ragged2e}
\usepackage{amssymb}
\usepackage{lipsum}
\usepackage[english]{babel}
\usepackage{mathtools}
\usepackage{braket}
\usepackage{graphicx}
\usepackage{wrapfig}
\usepackage{dsfont}
\usepackage{etoolbox}
\usepackage{comment}
\usepackage[colorlinks=true, linkcolor=blue, citecolor=cyan]{hyperref}
\usepackage{amsfonts}
\usepackage{soul}

\newcommand{\beq}[0]{\begin{equation}}
\newcommand{\eeq}[0]{\end{equation}}
\newcommand{\non}{\nonumber}
\def\be{\begin{equation}}
\def\ee{\end{equation}}
\def\bea{\begin{eqnarray}}
\def\eea{\end{eqnarray}}
\newcommand{\ba}{\begin{eqnarray}}
\newcommand{\ea}{\end{eqnarray}}

\begin{document}
\title{Precision bound and optimal control in periodically modulated continuous quantum thermal machines}
\author{Arpan Das}
\email{adas@fuw.edu.pl}
\affiliation{Institute of Physics, Faculty of Physics, Astronomy and Informatics,
Nicolaus Copernicus University, Grudzi{\k{ a}}dzka 5/7, 87-100 Toru{\'n}, Poland}

\author{Shishira Mahunta}
\email{shishiram@iiserbpr.ac.in}
\affiliation{Department of Physical Sciences,  Indian Institute of Science Education and Research Berhampur, Berhampur, 760010, India}

\author{Bijay Kumar Agarwalla}
\email{bijay@iiserpune.ac.in}
\affiliation{Department of Physics, Indian Institute of Science Education and Research Pune, Pune 411008, India}

\author{Victor Mukherjee}
\email{ mukherjeev@iiserbpr.ac.in}
\affiliation{Department of Physical Sciences,  Indian Institute of Science Education and Research Berhampur, Berhampur, 760010, India}

\begin{abstract}

We use Floquet formalism to study  fluctuations in periodically modulated continuous quantum thermal machines. We present a generic theory for such machines, followed by specific examples of sinusoidal, optimal, and circular modulations respectively. The thermodynamic uncertainty relations (TUR) hold for all modulations considered. Interestingly, in the case of sinusoidal modulation, the TUR ratio assumes a minimum at the heat engine to refrigerator transition point, while the Chopped Random Basis (CRAB) optimization protocol allows us to keep the ratio small for a wide range of modulation frequencies. Furthermore, our numerical analysis suggests that TUR can show signatures of heat engine to refrigerator transition, for more generic modulation schemes. We also study bounds in fluctuations in the efficiencies of such machines; our results indicate that fluctuations in efficiencies are bounded from above for a refrigerator, and from below for an engine. Overall, this study emphasizes the crucial role played by different modulation schemes in designing practical quantum thermal machines.

\end{abstract}

\date{\today}

\maketitle

\section{Introduction}
\label{intro}
The necessity of establishing a coherent framework of thermodynamics within the realm of quantum mechanics motivated the field of Quantum Thermodynamics \cite{kosloff13quantum,ghosh2018thermodynamic,vinjanampathy16quantum}. Such a framework would lead to a deeper understanding of the energetics in the quantum regime, which is crucial in the era of miniaturization of technologies. Moreover, along with the average thermodynamic quantities of interest, the fluctuations about the mean values are necessary for determining the quality of output of such miniature technologies.

In this regard, fluctuation theorems \cite{esposito-fluctuation-review, campisi-fluctuation-review, annurev-jarzynski, funo2018} are a remarkable achievement, providing exact mathematical expressions for arbitrary out of equilibrium scenario.  Beyond that, recently it has been shown that for out-of-equilibrium systems, there is a limitation to achieving arbitrary precision, which is broadly known as the Thermodynamic Uncertainty Relations (TUR) \cite{TUR1st,seifert-second,TUR-Gingrich,Pietzonka-1,horowitz-finite,TUR-tradeoff,tur-satis3}. Since its first analytical proof for time-continuous classical Markov jump process on a discrete set of states \cite{TUR-Gingrich}, it was subsequently proven for Langevin dynamics with continuous variables \cite{polettini-1,Gingrich_2017,Dechant_2018}. In the meantime, a plethora of studies have been carried out to broaden the regime of its applicability, to include  quantum systems. Beyond the standard framework, where the system relaxes to a unique time-independent steady state and the dynamics is time reversal symmetric \cite{Proesmans_2019}, the original TUR \cite{TUR1st, TUR-Gingrich} may not hold, thus requiring more general bounds. Specifically, for discrete-time Markov chains \cite{naoto-discrete,Proesmans_2017,discrete-3}, time dependent driving \cite{barato-clock,Proesmans_2017,Barato_2018, Koyuk_2018,sasa-2,holubec-cycling,TUR-Brandner,Barato_2019,koyuk-2020,benenti-2021,vu-under1}, underdamped Langevin equation \cite{vu-under2,udo-under,park-under,vu-under1}, ballistic transport in the presence of magnetic field \cite{TUR-Brandner,brander-transport}, with measurement and feedback control \cite{potts-feedback}, and arbitrary initial states \cite{ueda-tur-1}, modified TUR have been put forward. Extensions to the quantum scenario \cite{TURBijay2, TURDvira, bijay-exp, TURBijay1, large-deviation-quantum, goold-tur, hasegawa-tur1, hasegawa-tur2} are relatively less explored. In fact, additional complexity that arises due to the genuine quantum features, such as quantum coherence \cite{coherent-4, TURDvira, coherent-6, coherent-2, coherent-3, tur-satis, coherent-5, coherent-7} can potentially give rise to violation of the original TUR. 

Discovery of TUR have further provided a deep understanding in analysing the thermodynamics of precision for a thermal machine. Along with the power and efficiency to quantify the quality of an engine, the fluctuation around the power output 
enters into the trade-off relation following the original statement of TUR \cite{TUR1st,seifert-second,TUR-Gingrich}.  This was first shown for steady state classical heat engine connected simultaneously to two baths \cite{TUR-tradeoff}. As, the TUR does not always hold in its original form, in general such conclusion may not be true for all types of thermal machines. As an example, for cyclic heat engines this trade-off can be overcome \cite{holubec-cycling}. Nevertheless, following this work, generalizations have been put forward in both classical and quantum scenarios. For instance, quantum thermoelectric junctions \cite{TURDvira, coherent-4}, autonomous quantum thermal machines \cite{coherent-2, coherent-7}, periodic quantum thermal machines \cite{miller-slow,lee-otto-tur,sachhi-otto-tur, coherent-7} and information driven engines \cite{potts-feedback} have been studied. 

In this paper, we first study a periodically driven continuous quantum thermal machine \cite{klimovsky13minimal} where the working fluid consists of a minimal two-level system that can work both as an engine and  a refrigerator.  Employing the techniques of counting field statistics \cite{esposito-fluctuation-review} and the Floquet formalism \cite{floquet-1, floquet-2, floquet-3, floquet-4}, we provide analytic expressions for the currents and fluctuations. Very recently, a novel quantification of fluctuations has been introduced for heat engines \cite{ito2019universal, Bijay-machine-ss,Bijay-Otto}, defined as the ratio of  fluctuation of output work and fluctuation of heat input from the hot bath. Analogous definition has also been provided for refrigerator \cite{ Bijay-machine-ss,Bijay-Otto} defined as the ratio of the fluctuation of heat extracted from cold bath and fluctuation of input work. The mathematical form of this quantification is reminiscent of efficiency (co-efficient of performance for refrigerator) giving an essence of the fluctuation for relevant quantities. In general one can define $n-$th order for such quantities taking the ratio of $n-$th order central moments.
It was shown that such a quantity is bounded from both below and above for continuous autonomous thermal machines \cite{Bijay-machine-ss}. One advantage for using this quantity is that, as also shown previously for autonomous and discrete thermal machines, such quantity receives a universal upper bound that is solely dependent on the temperature ratio of the cold and hot heat baths.
We study the same quantity in the present context. We focus on cases where the qubit 
is driven by sinusoidal, optimal, and circular modulations. First demonstrating the validity of TUR for each scenario, we find that for such non-autonomous machines, while performing as an engine (refrigerator), the lower (upper) bound is always satisfied whereas we found that the upper (lower) bound is not satisfied at least for one model we considered.

 To illustrate further on TUR, in contrast to sinusoidal modulation, the optimized protocol obtained via Chopped Random Basis (CRAB) method allows for reduced TUR ratio for the input (heat current from the hot bath in the engine regime and power in the refrigerator regime) for a wide range of modulation frequencies, whereas the same may not be always true in the case of the output (power output in the engine regime and heat current extracted from the cold bath in the refrigerator regime). Furthermore, for such more generic modulation schemes, we observe that the TUR can show signatures of heat engine to refrigerator transition.
Finally, we focus on circular modulation with secular approximation \cite{secular-1, secular-2}. Combined with the Floquet analysis, we find that the evolution of the diagonal and the off-diagonal elements of the system density matrix are decoupled, leading to Pauli type master equation for the diagonal entries. Consequently, TUR is obeyed for heat currents in this model. In absence of secular approximation, \cite{secular-1, secular-2} evolution of diagonal and off-diagonal elements of the density matrix may not decouple, thereby leaving a possibility of violating the original TUR.
 
This paper is organized as follows. In Sec.~\ref{counting-description}, we introduce the techniques of counting field statistics to calculate the currents and fluctuations in periodically driven open quantum systems. In Sec.~\ref{section3}, we introduce the model of a minimal continuous quantum thermal machine; we derive the corresponding generalized master equation in Sec.~\ref{section3A}, and discuss the generating function and moments of heat currents and power in Sec.~\ref{section3B}. In Sec.~\ref{section3C} we present the results for  sinusoidal driving. We introduce the CRAB optimization protocol in Sec. \ref{secCRAB1}, and present the optimization results in Sec. \ref{secCRAB2}.  Section~\ref{section3D} deals with circular modulation. Finally, in Sec.~\ref{conclu}, we provide a summary of our main results. We provide certain technical details  in the appendices.

%%%%%%%%%%%%%%%%%%%%%%%%%%%%%%%%%%%%%%%%%%%%%

\section{Counting field statistics for driven open quantum systems}
\label{counting-description}
We write the total Hamiltonian for the system and the baths as
\begin{equation}
H(t)=H_S(t)+H_B+H_{SB}=H_0(t)+H_{SB},
\end{equation}
where, $H_B=H_{h}+H_{c}$ and, $H_{SB}=H_{Sh}+H_{Sc}=\mathcal{S}\otimes \mathcal{B}_h+\mathcal{S}\otimes \mathcal{B}_c$.
We are considering the situation where the two baths are continuously connected to the system. Here $H_h$ ($H_c)$ denotes the Hamiltonian of the hot (cold) bath; the operators $\mathcal{S}$ and $\mathcal{B}_j$ act on the system and the $j(= h, c)$th bath, respectively.  We choose the initial state as a product state i.e., $\rho(0)=\rho_S(0)\otimes \rho_B$, where $\rho_B=\rho_{h}\otimes \rho_{c}$ and the reservoirs are prepared in thermal states with respective Hamiltonians $H_h$, $H_c$ and inverse temperatures $\beta_h$, $\beta_c$, respectively.  The measured observables are the Hamiltonians $H_{h}$ and $H_{c}$. To get the probability distribution of this measurement, we introduce counting field $\chi_j$ $(j=h, c)$ to each reservoir. We introduce $\chi\equiv{\{\chi_h,\chi_c\}}$ to denote collectively both the counting variables. The generating function corresponding to the two-point measurement statistics is given by
\begin{equation}
\mathcal{G}(\chi,t)={\rm Tr}_{SB}[\rho(\chi,t)],
\end{equation}
where ${\rm Tr}_{SB}$ denotes tracing over both the bath and the system degrees of freedom.
The modified density matrix $\rho(\chi,t)$ is given as,
\begin{equation}
\label{EOM1}
\rho(\chi,t)=U(\chi,t)\rho(0)U^{\dagger}(-\chi,t),
\end{equation}
with,
\begin{equation}
U(\chi,t)=e^{-i(\chi_h H_{h}+\chi_c H_{c})/2}U(t)e^{i(\chi_h H_{h}+\chi_c H_{c})/2}
\end{equation}
being the counting field dressed evolution operator.  Here $U(t)$ is the unitary evolution operator generated by the total Hamiltonian $H(t)$.
Defining, $\rho_S(\chi,t)={\rm Tr}_B[\rho(\chi,t)]$, we get, 
\begin{equation}
\label{mom-gen}
\mathcal{G}(\chi,t)={\rm Tr}[\rho_S(\chi,t)].
\end{equation}
The generating function allows us to evaluate the statistics of energy transferred between system and each reservoir obtained from two point measurement scheme \cite{esposito-fluctuation-review}:
\begin{equation}
\label{moments}
\langle \Delta E_j^n\rangle=\frac{\partial^n}{\partial{(i\chi_j)}^n}\mathcal{G}(\chi,t)\mid_{\chi=0}, 
\end{equation}
The first order moment $\langle\Delta E_j\rangle$ is the heat transferred between the system and $j$-th reservoir. With the spectral decomposition of the bath Hamiltonian as
\begin{equation}
    H_j=\sum_m E_j^k\ket{k_j}\bra{k_j}\equiv\sum_k E_j^k P_j^k,
\end{equation}
$\langle\Delta E_j\rangle$ is defined below, where two projective measurements (the projectors $\{P_j^k\}$ are defined above) are done on $H_j$ at the beginning and at time $t$.
\begin{equation}
    \langle \Delta E_j \rangle=\sum_{m,n}p_j^m p_j^{nm}(E_j^n-E_j^m)
\end{equation}
where,
\begin{equation}
    p_j^{nm}={\rm Tr}[P_j^n U(t)P_j^m\rho(0)P_j^m U^{\dagger}(t)P_j^n]
    \label{def_prob}
\end{equation}
and $p_j^m={\rm Tr}[\rho(0)P_j^m]$ is the probability to measure $E_j^m$ at $t=0$.
In Eq. (\ref{def_prob}) and in the definition of $p_j^m$, the projector $P_j^k$ is understood as $\mathds{1}\otimes P_j^k$.
Now Eq. (\ref{moments}) in turn results in the  mean heat current given by,
\begin{equation}
\label{heat-current}
\langle J_j(t)\rangle=\frac{d}{dt}\langle \Delta E_j\rangle=-i\frac{d}{dt}\frac{\partial}{\partial\chi_j}\mathcal{G}(\chi,t)\mid_{\chi=0}.
\end{equation}
As per our convention, positive (negative) sign implies current is entering (leaving) the system. 

We now proceed to study the dynamics of  $\rho(\chi,t)$. Following Eq.~(\ref{EOM1}), the evolution of this modified density matrix is given as \cite{esposito-fluctuation-review},
\begin{equation}
\partial_t\rho(\chi,t)=-i[H(\chi,t)\rho(\chi,t)-\rho(\chi,t)H(-\chi,t)],
\label{rhoeq1}
\end{equation}
where, $H(\chi,t)=e^{-i(\chi_h H_h+\chi_c H_{c})/2}H(t)e^{i(\chi_h H_h+\chi_c H_{c})/2}$.
In the interaction picture one gets (the operators are labelled by tilde)
\begin{equation}
\tilde{\rho}(\chi,t)=U_0\rho(\chi,t)U^{\dagger}_0, 
\end{equation}
where $U_0$ is the unitary operator generated by the Hamiltonian $H_0(t)=H_S(t)+H_B$; the interaction picture Hamiltonian is given by
\begin{equation}
\tilde{H}_I(\chi,t)=U_0 H_{SB}(\chi,t)U_0^\dagger=\mathcal{\tilde{S}}(t)\otimes (\mathcal{\tilde{B}}_h(\chi_h,t)+\mathcal{\tilde{B}}_c(\chi_c,t))
\end{equation} 
In the interaction picture, the equation of motion \eqref{rhoeq1} now becomes
\begin{equation}
\partial_t\tilde{\rho}(\chi,t)=-i[\tilde{H}_I(\chi,t)\tilde{\rho}(\chi,t)-\tilde{\rho}(\chi,t)\tilde{H}_I(-\chi,t)],
\end{equation}
Next, considering the weak coupling assumption and performing the standard Born-Markov approximation, we arrive at the following master equation
\begin{align}
\nonumber
\partial_t\tilde{\rho}_S(\chi,t)&=-\int_0^{\infty}d\tau{\rm Tr}_B[\tilde{H}_I(\chi,t)\tilde{H}_I(\chi,t-\tau)\tilde{\rho}_S(\chi,t)\rho_B\\
\nonumber
&-\tilde{H}_I(\chi,t)\tilde{\rho}_S(\chi,t)\rho_B\tilde{H}_I(-\chi,t-\tau)\\
\nonumber
&-\tilde{H}_I(\chi,t-\tau)\tilde{\rho}_S(\chi,t)\rho_B\tilde{H}_I(-\chi,t)\\
\label{parent-master1}
&+\tilde{\rho}_S(\chi,t)\rho_B\tilde{H}_I(-\chi,t-\tau)\tilde{H}_I(-\chi,t)],
\end{align}
where we have used ${\rm Tr}_B[\tilde{H}_I(\chi,t)\rho_B]=0$ \cite{breuer02}.
The first term on the r.h.s of Eq. (\ref{parent-master1}) can be written as
\begin{align}
\label{parent-eq}
\sum_{j=h,c}\int_0^\infty d\tau \, \mathcal{\tilde S}(t)\mathcal{\tilde S}(t-\tau)\tilde{\rho}_S(\chi,t)\Phi_j(\tau),
\end{align}
where we have ${\rm Tr}_{\tilde B_j}[\mathcal{\tilde B}_j(\chi,t_1)\mathcal{\tilde B}_j(\eta,t_2)\rho_j]={\rm Tr}_{\tilde  B_j}[\mathcal{\tilde  B}_j(\chi-\eta,t_1-t_2)\mathcal{\tilde  B}_j\rho_j]\equiv\Phi_j(\chi-\eta,t_1-t_2)$, and $\Phi_j(0,t)=\Phi_j(t)$. Also, $\mathcal{\tilde{S}}(t)$ is the system operator (in the interaction picture) coming from the interaction Hamiltonian $H_{SB}$.
Similarly, evaluating the other terms, we finally have
\begin{align}
\label{Gen-master-eq}
\nonumber
\partial_t\tilde{\rho}_S(\chi,t)&=-\sum_{j=h,c}\int_0^\infty d\tau [\mathcal{\tilde S}(t)\mathcal{\tilde  S}(t-\tau)\tilde{\rho}_S(\chi,t)\Phi_j(\tau)\\
\nonumber
&-\mathcal{\tilde  S}(t)\tilde{\rho}_S(\chi,t)\mathcal{\tilde S}(t-\tau)\Phi_j(-2\chi,-\tau)\\
\nonumber
&-\mathcal{\tilde  S}(t-\tau)\tilde{\rho}_S(\chi,t)\mathcal{\tilde S}(t)\Phi_j(-2\chi,\tau)\\
&+\tilde{\rho}_S(\chi,t)\mathcal{\tilde S}(t-\tau)\mathcal{\tilde S}(t)\Phi_j(-\tau)].
\end{align}
Equation \eqref{Gen-master-eq} is the generalized master equation for a generic quantum system in presence of arbitrary modulation, the solution of which will provide the required generating function. For later convenience, we introduce the Fourier transform of the correlation functions, 
\begin{equation}
\Phi_j(-2\chi,\tau)=\frac{1}{2\pi}\int_{-\infty}^{\infty}d\nu e^{-i\nu(\chi+\tau)}G_j(\nu).
\end{equation}

%%%%%%%%%%%%%%%%%%%%%%%%%%%%%%%%%%%%%%%%%%%%%

\section{A minimal continuous quantum thermal machine}
\label{section3}
In this section, we focus on the specific case of a thermal machine modelled by a minimal (two-level system) working medium in presence of a periodic modulation \cite{klimovsky13minimal}. 
%{\color{blue}The authors in the Ref. \cite{klimovsky13minimal} called the model minimal in the sense that it is using a minimal model (qubit) of the WM to have a quantum mechanical Hamiltonian which can work both as engine or refrigerator and is adaptable to available baths and temperature. We have}
\ba
\label{model-hamil}
H_S(t) = H_S(t+T);~~~H_S(t) = \frac{1}{2}\omega(t)\sigma_z,
\label{Eq:hamil}
\ea
with period $T=2\pi/\Delta$. Additionally, we consider $\mathcal{S}=\sigma_x$. One can represent $\mathcal{\tilde S}(t)=\sigma_x(t)=U^\dagger_S(t) \sigma_x U_S(t)$ in the Floquet basis as \cite{alicki14quantum}
\ba
\label{x-expr}
\tilde {\sigma}_x(t)&=&\sum_{q\in \mathcal{Z}}(\eta(q)e^{-i(\omega_0+q\Delta)t}\sigma^{-}+\eta^*(q)e^{i(\omega_0+q\Delta)t}\sigma^+),\non\\
\eta(q) &=& \frac{1}{T} \int_0^T \exp\left(i \int^t_0 \left(\omega(s) - \omega_0 \right) ds\right) e^{-iq \Delta t} dt.
\ea 
To justify the above equation we refer to the Floquet theory discussed in detail in Appendix \ref{appflq}.

We note that the resonance condition $\Delta = \omega_0 = \frac{1}{T}\int_0^T\, dt \, \omega(t)$ may result in effective squeezing, as reported in Ref. \cite{squeez-1}. Furthermore, experimental constraints may impose a limitation on the maximum frequencies of modulation in any setup. Consequently, in this study we restrict ourselves to low-frequency modulation only, given by $\Delta < \omega_0$. 

We note that one can consider an autonomous quantum thermal machine as well, which can operate even in the absence of any external periodic modulation \cite{niedenzu19concepts}. However, such machines are outside the scope of this present work. 
%%%%%%%%%%%%%%%%%%%%%%%%%%%%%%%%%%%%%%%%%%%%%

\subsection{Generalized quantum master equation with counting fields}
\label{section3A}
With the above expansion of $\tilde {\sigma}_x(t)$ (see Eq. \eqref{x-expr}), we perform a secular approximation for Eq. (\ref{Gen-master-eq});
neglecting the terms with $q\neq q'$, the first term of the expression Eq. (\ref{Gen-master-eq}) becomes
\begin{align}
\nonumber
\frac{1}{2}\sum_{q,j} {|\eta(q)|}^2&(G_j(-\omega_0-q\Delta)\sigma^-\sigma^+\tilde{\rho}_S(\chi,t)\\
&+G_j(\omega_0+q\Delta)\sigma^+\sigma^-\tilde{\rho}_S(\chi,t)),
\end{align} 
where $j=\{h,c\}$. In the above we have neglected the Cauchy principal value and used,
 \begin{equation}
 \int_0^{\infty} d\tau e^{i\tau(y-x)}=\pi\delta(y-x).
 \end{equation}
Following similar steps and writing for both hot and cold baths, we finally obtain,
 \begin{align}
 \label{master-chi}
 \nonumber
 &\partial_t\tilde{\rho}_S(\chi,t)=\sum_{q,j=\{h,c\}}\mathcal{L}^{q}_{i}(\chi_j,t)[\tilde{\rho}_S(\chi,t)]=&\\
 \nonumber
 \sum_{q,j=\{h,c\}}\frac{P_q}{2}&\Big(G_j(\omega_0+q\Delta)[2e^{-i(\omega_0+q\Delta)\chi_j}\sigma^-\tilde{\rho}_S(\chi,t)\sigma^+\\
 \nonumber
 &-\sigma^+\sigma^-\tilde{\rho}_S(\chi,t)-\tilde{\rho}_S(\chi,t)\sigma^+\sigma^-]&\\
 \nonumber
 &+G_j(-\omega_0-q\Delta)[2e^{i(\omega_0+q\Delta)\chi_j}\sigma^+\tilde{\rho}_S(\chi,t)\sigma^-\\
 &-\sigma^-\sigma^+\tilde{\rho}_S(\chi,t)-\tilde{\rho}_S(\chi,t)\sigma^-\sigma^+]\Big),
 \end{align}
where 
 $P_q={|\eta(q)|}^2$ is the weight of the $q-$th Floquet mode. For $\chi=0$, we get back the original master equation without the counting field $\chi$ \cite{esposito-fluctuation-review}. We note that here for simplicity we have assumed $\omega_0 + q\Delta$ is positive for all the significant Floquet modes, which can be the case for modulations with $P_q \to 0$ for large $|q|$. A more general case with positive as well as negative $\omega_0 + q\Delta$ is discussed in Appendix \ref{appec}.

 %%%%%%%%%%%%%%%%%%%%%%%%%%%%%%%%%%%%%%%%%%%%%

 \subsection{Generating function, mean currents and fluctuations}
 \label{section3B}
With the above generalized master equation \eqref{master-chi} in hand, we now compute the mean and fluctuations of the heat currents and the output power. Importantly,  for the master equation in Eq.~(\ref{master-chi}), the evolution of the diagonal and off-diagonal elements are decoupled. As a result, it is enough to consider only the evolution of  the diagonal entries of $\tilde{\rho}_S(\chi,t)$, where we have noted that the generating function is given by the trace of $\tilde{\rho}_S(\chi,t)$ (see Eq. (\ref{mom-gen})).
 In the energy eigenbasis, the  time evolution of the diagonal entries ${\tilde{\rho}}_{00}(\chi,t)$ and ${\tilde{\rho}}_{11}(\chi,t)$ can be expressed as
\begin{align}
\label{vector-form}
%\nonumber
&\begin{pmatrix}
\dot{\tilde{\rho}}_{00}(\chi,t)\\
\dot{\tilde{\rho}}_{11}(\chi,t)
\end{pmatrix}=\mathcal{L}(\chi)\begin{pmatrix}
\tilde{{\rho}}_{00}(\chi,t)\\
\tilde{{\rho}}_{11}(\chi,t)
\end{pmatrix},
\end{align}
where the elements of the matrix 
\ba
\mathcal{L}(\chi) = \left[ \begin{array}{cc} l_{00} & l_{01}^{\chi} \\
l_{10}^{\chi} & l_{11} \end{array} \right]
\ea
are 
\begin{align}
&l_{00}=-\sum_{q,j} P_q G_j(\omega_0+q\Delta)\\
 &l_{01}^{\chi}= \sum_{q,j} P_q G_j(-\omega_0-q\Delta)e^{i(\omega_0+q\Delta)\chi_j}\\
&l_{10}^{\chi}=\sum_{q,j} P_q G_j(\omega_0+q\Delta)e^{-i(\omega_0+q\Delta)\chi_j}\\
  &l_{11}= -\sum_{q,j} P_q G_j(-\omega_0-q\Delta)
\end{align}
We use the moment generating function Eq. (\ref{mom-gen}) to define the cumulant generating function as,
\begin{equation}
\mathcal{C}(\chi,t) \equiv \log \mathcal{G}=\log {\rm Tr}[\tilde{\rho}_S(\chi,t)],
\end{equation} 
which directly gives us the mean, variance and the higher order cumulants.
The steady state is reached in the limit of  long times, when 
the cumulant generating function is dominated by the eigenvalue $\lambda(\chi)$ of  $\mathcal{L}(\chi)$ with the largest real part.  Therefore one can write \cite{schaller2014open},
\begin{equation}
\label{cumulantss}
\lim_{t\rightarrow \infty}\mathcal{C}(\chi,t)\approx \lambda(\chi)t,
\end{equation}
where, 
\begin{equation}
\lambda(\chi)=\frac{1}{2}(l_{00}+l_{11}) + \frac{1}{2}\sqrt{{(l_{00}+l_{11})}^2-4(l_{00}l_{11}-l_{01}^{\chi}l_{10}^{\chi})}.
\end{equation}

This in turn results in the mean current in the steady to be given by,
\begin{align}
\label{formula1}
& \langle J_j\rangle= \left.\lim_{t\rightarrow \infty}\frac{d}{dt}\frac{\partial}{\partial(i\chi_j)}\mathcal{C}(\chi,t)\right\vert_{\chi=0}=\left.\frac{\partial \lambda(\chi)}{\partial(i\chi_j)}\right\vert_{\chi=0,}\\
&=\sum_q \frac{P_q (\omega_0+q\Delta)}{w+1}G_j(\omega_0+q\Delta)[e^{-\beta_j(\omega_0+q\Delta)}-w].
\end{align}

Here, $w$ is the ratio of the diagonal entries of $\tilde{\rho}_S (0, t)$ in the steady state, evaluated setting $\chi=0$ in the Eq. (\ref{vector-form}) and can be obtained straightforwardly as, 
\begin{equation}
\label{p11p22ss1}
\frac{p_1^{ss}}{p_2^{ss}}\equiv w=\frac{\sum_{q,j}P_qG_j(\omega_0+q\Delta)e^{-\beta_j(\omega_0+q\Delta)}}{\sum_{q,j}P_qG_j(\omega_0+q\Delta)}=\frac{l_{11}}{l_{00}}.
\end{equation}
To arrive at this expression we have used the Kubo-Martin-Schwinger (KMS) boundary condition \cite{alicki2007quantum}, 
\begin{equation}
G_j(-\nu)=e^{-\nu}G_j(\nu);~~~\nu > 0.
\label{KMScondn}
\end{equation}
and assumed $\left(\omega_0+q\Delta\right) > 0$ for simplicity.
 The KMS condition above is the direct consequence of the fact that the bath states are in thermal equilibrium with inverse temperature $\beta_j$. For setups with both positive as well as negative $\left(\omega_0+q\Delta\right)$, one can consider the same KMS condition \eqref{KMScondn} to do the analysis, as discussed in Appendix \ref{appec}. 
Similarly, the current fluctuation is given as,
\begin{align}
\label{formula2}
& {\rm var}(J_j)=\left.\lim_{t\rightarrow \infty}\frac{d}{dt}\frac{\partial^2}{\partial{(i\chi_j)}^2}\mathcal{C}(\chi,t)\right\vert_{\chi=0}=\left.\frac{\partial^2\lambda(\chi)}{\partial(i\chi_j)^2}\right\vert_{\chi=0}, \nonumber \\
\end{align}
which upon simplification leads to the following expression:
\begin{widetext}
\begin{align}
\nonumber
&{\rm var}(J_j)=\sum_q \frac{P_q {(\omega_0+q\Delta)}^2}{w+1}G_j(\omega_0+q\Delta)[e^{-\beta_j(\omega_0+q\Delta)}+w]-\frac{2{\langle J_j\rangle}^2}{\sum_{q,j}P_qG_j(\omega_0+q\Delta)[e^{-\beta_j(\omega_0+q\Delta)}+1]}\\
&-\frac{2}{\sum_{q,j}P_qG_j(\omega_0+q\Delta)[e^{-\beta_j(\omega_0+q\Delta)}+1]}\Big[\sum_{q',q''}P_{q'} P_{q''}e^{-\beta_j(\omega_0+q'\Delta)}(\omega_0+q'\Delta)(\omega_0+q''\Delta)G_j(\omega_0+q'\Delta)G_j(\omega_0+q''\Delta)\Big].
\end{align}
\end{widetext}
One can use the above results to evaluate the average entropy production rate
\begin{equation}
\langle \dot{S}\rangle =-\beta_h \langle J_h\rangle-\beta_c \langle J_c \rangle,
\label{eqen}
\end{equation}
and average output power
\begin{equation}
\langle \mathcal{P}\rangle=-\langle J_h\rangle-\langle J_c \rangle,
\end{equation}
and fluctuations (variance) in power
\begin{equation}
{\rm var}(\mathcal{P})={\rm var}(J_h)+{\rm var} (J_c)+2~ {\rm Cov}(J_h,J_c).
\end{equation}
Here the covariance term is given by ${\rm Cov}(J_h,J_c)=\langle J_hJ_c\rangle-\langle J_h\rangle \langle J_c\rangle$, where,
\begin{widetext}
\ba
&\langle J_hJ_c\rangle=\left.\frac{\partial^2 \lambda(\chi)}{\partial (i\chi_c)\partial(i\chi_h)}\right\vert_{\chi=0}
=\frac{1}{l_{00}+l_{11}}\Big(\left.\frac{\partial l_{10}^{\chi}}{\partial(i\chi_c)}\frac{\partial l_{01}^{\chi}}{\partial(i\chi_h)}+\frac{\partial l_{01}^{\chi}}{\partial(i\chi_c)}\frac{\partial l_{10}^{\chi}}{\partial(i\chi_h)}\Big)\right\vert_{\chi_j=0}-\frac{2}{l_{00}+l_{11}}\langle J_h\rangle \langle J_c \rangle;\non\\
\nonumber
&\Big(\left.\frac{\partial l_{10}^{\chi}}{\partial(i\chi_c)}\frac{\partial l_{01}^{\chi}}{\partial(i\chi_h)}+\frac{\partial l_{01}^{\chi}}{\partial(i\chi_c)}\frac{\partial l_{10}^{\chi}}{\partial(i\chi_h)}\Big)\right\vert_{\chi_j=0}=-\sum_{q',q''}P_{q'} P_{q''}e^{-\beta_h(\omega_0+q'\Delta)}(\omega_0+q'\Delta)(\omega_0+q''\Delta)G_h(\omega_0+q'\Delta)G_c(\omega_0+q''\Delta)\\
&-\sum_{q',q''}P_{q'} P_{q''}e^{-\beta_c(\omega_0+q''\Delta)}(\omega_0+q'\Delta)(\omega_0+q''\Delta)G_h(\omega_0+q'\Delta)G_c(\omega_0+q''\Delta).
\label{eqcovJ}
\ea
\end{widetext}
Next, we demonstrate the validity of TUR for the heat currents and power for the specific examples of sinusoidal and circular modulations. In addition, we also study the bounds on the ratio of fluctuations for the currents.

%%%%%%%%%%%%%%%%%%%%%%%%%%%%%%%%%%%%%%%%%%%%%
\subsection{Sinusoidal modulation}
\label{section3C}
In this section we focus on the specific case of sinusoidal modulation: 
\ba
\omega(t)=\omega_0+\lambda \, \Delta\, \sin(\Delta t).
\label{eqsin}
\ea 
Here we assume a weak modulation, quantified by  $0\leq \lambda \leq 1$.
This assumption allows us to consider only the harmonics $q=0,\pm 1$, with,
\ba
P_0\approx 1-\frac{\lambda^2}{2}, ~ P_{\pm 1} \approx \frac{\lambda^2}{4},
\label{eqP}
\ea
with the higher order harmonics $P_{q} \approx 0$ for $|q| > 1$ in the limit of small $\lambda$.  Additionally, we consider the bath spectral functions such that 
\begin{equation}
G_c(\omega)\approx 0~ \text{for}~\omega\geq \omega_0, ~~ G_h(\omega)\approx 0~ \text{for}~\omega\leq \omega_0.
\label{eqG}
\end{equation}
The above choice  of spectral separation of the baths in Eq.~\eqref{eqG} allows the hot (cold) bath to interact with the WM only at higher (lower) energies, which is crucial for the operation of the quantum thermal machine as a heat engine or a refrigerator \cite{klimovsky13minimal, klimovsky15thermodynamics}. This is reminiscent of an Otto cycle, where the WM interacts with the hot (cold) bath only at higher (lower) frequencies \cite{kosloff17the}.
%The above choice  of spectral functions in Eq.~\eqref{eqG} allows us to implement spectral separation of the baths, which is crucial for the operation of the quantum thermal machine as a heat engine or a refrigerator \cite{klimovsky13minimal, klimovsky15thermodynamics}.
One can use Eqs. \eqref{formula1} - \eqref{eqcovJ} and Eq. \eqref{eqP} to arrive at the mean heat currents, power and their fluctuations (see Appendix \ref{appena}).

With the above setup, one can show that there exists a critical modulation frequency $\Delta \equiv \Delta_{cr} = \omega_0 \left(T_{ h} - T_{ c}\right)/\left(T_{ h} + T_{ c}\right)$, such that the machine works as a heat engine ($\langle J_h \rangle >0$, $\langle  J_c \rangle<0$, $\langle  \mathcal{P} \rangle<0$) for $\Delta<\Delta_{cr}$ and as a refrigerator ($\langle  J_h \rangle<0$, $\langle  J_c \rangle>0$, $\langle  \mathcal{P} \rangle>0$) for $\Delta >\Delta_{cr}$. The power output vanishes while the efficiency approaches the Carnot limit for $\Delta \to \Delta_{\rm cr}$ \cite{klimovsky13minimal}.
%(see Fig. \ref{allcurrents-1}) and Ref. \cite{klimovsky13minimal}).

We now focus on fluctuations in the operation of the thermal machine described above. To this end, we choose Lorentzian forms for the spectral functions of heat baths described by 
\ba
G_{h}(\omega\geq 0) &=& \frac{\gamma_0\Gamma^2 \Theta\left(\omega - \omega_0 -\epsilon \right)}{{(\omega-\omega_0-\delta)}^2+\Gamma^2};\non\\
G_{c}(\omega\geq 0) &=& \frac{\gamma_0\Gamma^2 \Theta\left(\omega_0 - \epsilon - \omega \right)}{{(\omega-\omega_0-\delta)}^2+\Gamma^2};\non\\
G_j(\omega<0) &=& G_j(\omega\geq 0)e^{-\beta_h\omega};~~~j=\{h,c\}
\label{GLorenz}
\ea
Here, $\gamma_0$ is the coupling strength, $\Gamma$ is the width of the spectrum, $\Theta$ denotes the Heaviside function,  $\epsilon$ is an infinitesimally small positive constant which ensures that $G_{ h}(0 \leq \omega < \omega_0) =  G_{ c}(\omega> \omega_0) = 0$ (see Eq. \eqref{eqG}), and $\delta > 0$ is an energy shift, such that $G_h(\omega)$ assumes maximum value at $\omega = \omega_0 + \delta$. 

A consistently performing thermal machine demands low noise-to-signal ratio ${\rm var}\left(\mathcal{J}\right)/\langle \mathcal{J}\rangle^2$, for $\mathcal{J}\in\{J_h,J_c, \mathcal{P}\}$. On the other hand, a low noise to signal ratio may imply a high entropy production rate $\langle \dot{S} \rangle $, which  in turn can result in low efficiency \cite{sachhi-otto-tur}. Consequently, one trade off relation for precision and cost of a thermal machine is the TUR:
\begin{equation}
\mathcal{R}_j = \langle \dot{S}  \rangle \frac{{\rm var}(\mathcal{J})}{{\mathcal{\langle J\rangle}^2}},
\label{eqTUR}
\end{equation}
where $j = h, c$ or $\mathcal{P}$, for $\mathcal{J} = J_h,J_c$ and   $\mathcal{P}$ respectively.
For the quantum thermal machine discussed here, the occupation probabilities $p_0= \rho_{00}$ and $p_1=\rho_{11}$ of the energy eigenstates of $H_S$ follow a Pauli-type master equation with time-independent coefficients, as shown in Eq.~(\ref{vector-form}) (setting $\chi=0$). Consequently, one can expect that with the KMS condition, the conventional TUR relation, 
\ba
\mathcal{R}_j \geq 2
\label{eqTURineq}
\ea
is satisfied \cite{Proesmans_2019,tur-satis2, tur-satis3, tur-satis}. 
\begin{figure}
\begin{center}
\includegraphics[width=0.9\columnwidth]{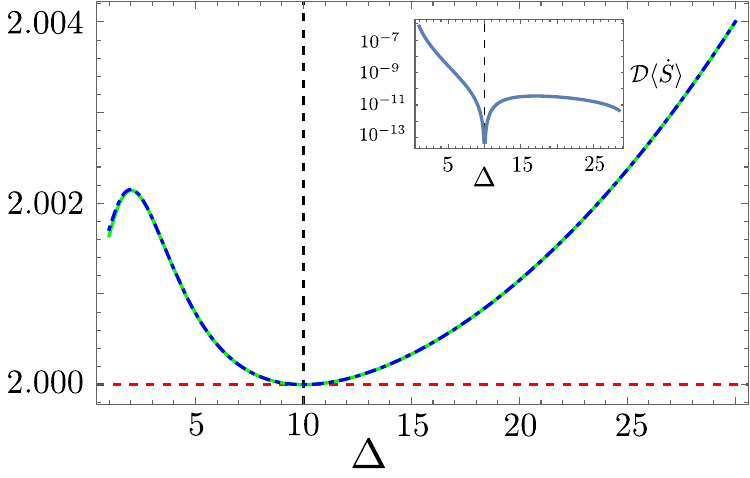}
\caption{Plot of TUR ratio for heat current $J_h$ (green solid line) and power $\mathcal{P}$ (blue dashed-dotted line) with $\Delta$, for sinusoidal modulation. Here, $\omega_0=30$, $\beta_h=0.005$, $\beta_c=0.01$, $\lambda=0.02$, $\Gamma=0.2$, $\delta=3$, and $\gamma_0=1$. Critical modulation frequency $\Delta_{cr}=10$, separating the heat engine and the refrigerator regimes are denoted by the vertical black dashed line. The red horizontal dashed line represents the value of the TUR ratio's lower limit (2). The TUR value saturates to the lower limit at the critical frequency $\Delta_{cr}$. Inset: We plot the difference between the TUR ratio for the power $\mathcal{P}$ and the heat current $J_h$, as a function of $\Delta$.}
\label{tur-1}
\end{center}
\end{figure}
We  validate the above inequality \eqref{eqTURineq} numerically,  as shown in Fig. \ref{tur-1}; here we display the TUR ratio for hot and power currents as a function of $\Delta$. For both cases, the TUR ratio is always lower bounded by the value 2. Note that, for  
$\Delta <\Delta_{cr}$, the machine works as an engine and for $\Delta > \Delta_{cr}$ the machine works as a refrigerator. As $\Delta$ approaches $\Delta_{cr}$ the TUR value tends to saturate to 2 as at this crossover point all the currents vanish. 
Furthermore, the expressions for the mean and variance of the currents imply (see Appendix \ref{appena}),
\begin{align}
&\frac{{{\rm {var}}(J_h)}}{\langle J_h\rangle^2}=\frac{{{\rm {var}}(J_c)}}{\langle J_c\rangle^2},\label{diff-fluc1}\\
\mathcal{D}:=&\frac{{{\rm {var}}(\mathcal{P})}}{\langle \mathcal{P}\rangle^2}-\frac{{{\rm {var}}(J_h)}}{\langle J_h \rangle^2}=\frac{1}{2}\left(\frac{\omega_0^2}{\Delta^2}-1\right).
\label{diff-fluc2}
\end{align}
Interestingly, the relative fluctuations for both hot and cold currents are the same. As a result, the TUR ratio for both the currents will also be the same and will always be lower bounded by the value 2, i.e., 
\begin{align}
& \langle \dot{S} \rangle \frac{{{\rm {var}}(J_h)}}{\langle J_h\rangle^2}= \langle \dot{S} \rangle \frac{{{\rm {var}}(J_c)}}{\langle J_c\rangle^2} \geq 2.
\label{ineq1}
\end{align}
The second relation in Eq.~(\ref{diff-fluc2}) leads to some important consequences: In the absence of strong modulation ($\Delta < \omega_0$), as $\langle \dot{S}\rangle \geq 0$, one immediately obtains $\mathcal{D}\langle \dot{S}\rangle\geq 0$, implying,
\begin{align}
& \langle \dot{S} \rangle \frac{{{\rm {var}}(\mathcal{P})}}{\langle \mathcal{P}\rangle^2}\geq \langle \dot{S} \rangle \frac{{{\rm {var}}(J_h)}}{\langle J_h\rangle^2} \geq 2.
\label{ineq2}
\end{align}
The above inequalities \eqref{ineq1} and \eqref{ineq2} show that reduction in the entropy production rate $\langle \dot{S} \rangle$ comes at the cost of higher noise to signal ratios of the heat currents and power, thus signifying a trade-off between the two.
The inset of Fig.~\ref{tur-1} shows  the quantity $\mathcal{D}\langle\dot{S}\rangle$ in log scale. 
At the crossover point ($\Delta=\Delta_{\rm cr}$), $\mathcal{D}$ is finite (Eq. (\ref{diff-fluc2})), whereas $\langle\dot{S}\rangle$ is zero. Consequently, this gives rise to $\mathcal{D}\langle\dot{S}\rangle$ vanishing at $\Delta=\Delta_{\rm cr}=10$, as also verified numerically in the inset of Fig.~\ref{tur-1}.
\begin{figure}
\begin{center}
\includegraphics[width=0.46\textwidth]{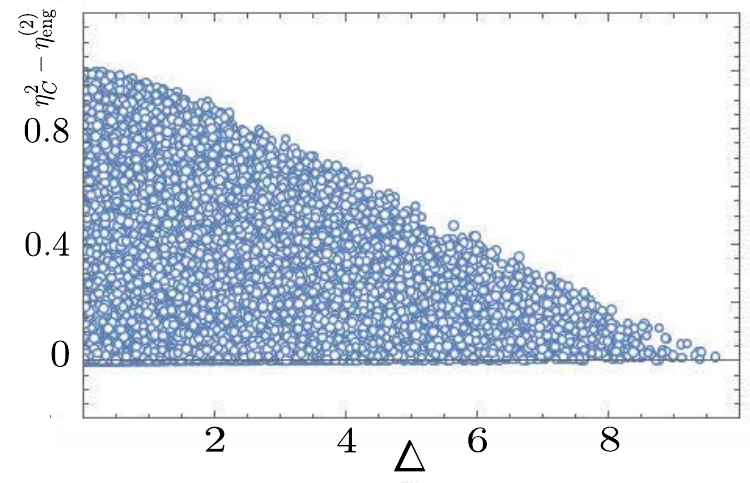}
\caption{ Scatter plot of $\eta_C^2-\eta^{(2)}_{\rm eng}$ in the engine regime, for sinusoidal modulation. We have considered $10^5$ sets of values for $\omega_0$, $\beta_h$, $\beta_c$, and $\Delta$ in the range $(0,10)$, with the constraints $\beta_h<\beta_c$, and $\Delta<\Delta_{\rm cr}$. For the values of $\Delta$ along the $x$-axis , we plot the corresponding values of $\eta_C^2-\eta^{(2)}_{\rm eng}$ along the $y$-axis.
Here, $\lambda=0.02$, $\Gamma=0.2$, $\delta=3$, and $\gamma_0=1$.}
\label{upper-engine}
\end{center}
\end{figure}

We next discuss another recently obtained bound on the ratio of fluctuations of currents and power; following Ref.  \cite{Bijay-machine-ss}, we define the following quantities:
\begin{align}
&\eta^{(2)}_{\rm eng}=\frac{{{\rm {var}}(\mathcal{P})}}{{\rm var}(J_h)}, ~~\text{for engine};\non\\
&\eta^{(2)}_{\rm ref}=\frac{{{\rm{var}}(J_c)}}{{\rm var}(\mathcal{P})}, ~~\text{for refrigerator}.
\label{eqetafluc}
\end{align}
Based on Onsager's reciprocity relations in the linear response regime for autonomous continuous machines, it was recently shown that $\eta^{(2)}$ is bounded from both above as well as below, as \cite{Bijay-machine-ss}
\begin{align}
&\langle \eta\rangle_{\rm eng} ^2\leq \eta^{(2)}_{\rm eng}\leq \eta_C^2, ~~\text{for engine}; \non \\
&\langle\eta\rangle_{\rm ref}^2\leq \eta^{(2)}_{\rm ref}\leq {\left(\frac{1-\eta_C}{\eta_C}\right)}^2=\eta_R^2, ~~\text{for refrigerator}.
\label{eq:effref}
\end{align}
Here $\langle\eta\rangle_{\rm eng}=\mathcal{\langle P\rangle}/\langle J\rangle_h$ and $\langle\eta\rangle_{\rm ref}=\langle J\rangle_c/\mathcal{\langle P\rangle}$ denote the efficiency of the engine and the coefficient of performance for the refrigerator, respectively. $\eta_C=1-\frac{\beta_h}{\beta_c}$ is the Carnot efficiency. Knowledge about such bounds in the present context is crucial for designing optimal non-autonomous continuous quantum thermal machines beyond linear response. We therefore now assess the validity of these bounds for our periodically driven continuous quantum thermal machine setup. 
\begin{figure}
\begin{center}
\includegraphics[width=0.9\columnwidth]{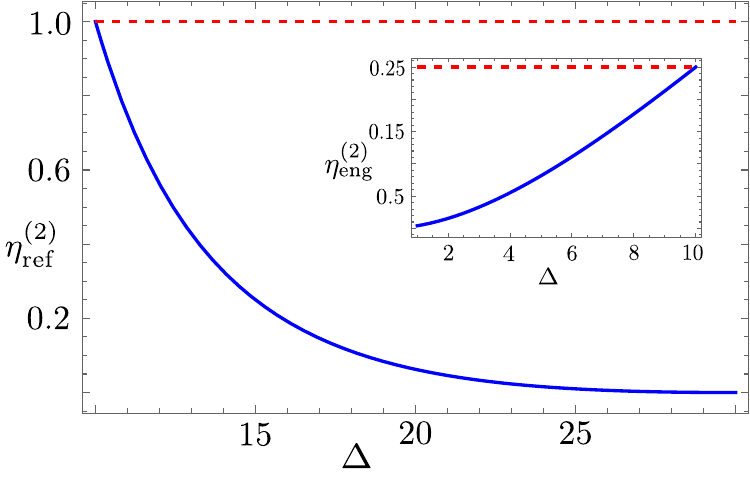}
\caption{Plot of $\eta_{\rm ref}^{(2)}$ with $\Delta$, for sinusoidal modulation. Here, $\omega_0=30$, $\beta_h=0.005$, $\beta_c=0.01$, $\lambda=0.02$, $\Gamma=0.2$, $\delta=3$, and $\gamma_0=1$. Critical modulation frequency $\Delta_{cr}=10$. The horizontal red dashed line represents $\eta_R^2=1$. Inset: We plot $\eta_{\rm eng}^{(2)}$ with $\Delta$ for the same parameter values. The red dashed horizontal line represents $\eta_C^2=0.25$.}
\label{upper-saturate}
\end{center}
\end{figure}
Following the above expressions in Eq. (\ref{diff-fluc1}) and Eq. (\ref{diff-fluc2}) for relative fluctuations in the engine regime, we arrive at the following relations:
\begin{align}
&\eta^{(2)}_{\rm eng}=\frac{{{\rm {var}}(\mathcal{P})}}{{\rm var}(J_h)}>\frac{\langle \mathcal{P} \rangle ^2}{\langle J_h \rangle^2}=\langle\eta\rangle_{\rm eng}^2,
\label{eng-low-bound}
\end{align}
 therefore implying that the lower bound gets respected. Note that, as in this case, the equality in the lower bound is never achieved in the engine regime $\Delta < \omega_0$.  Next, in the refrigerator regime, we find,
\begin{align}
\label{upper-bound-refrigerator}
&\eta^{(2)}_{\rm ref}=\frac{{{\rm {var}}(J_c)}}{{\rm var}(\mathcal{P})}< \frac{\langle J_c \rangle^2}{\langle \mathcal{P} \rangle^2}=\langle\eta\rangle_{\rm ref}^2 \leq \eta_R^2,
\end{align}
thus signifying the violation of the lower bound but always validating the upper bound.  Note that, for finite-time discrete quantum Otto cycle, a setup that is in stark contrast to the present situation, similar observations were pointed out \cite{Bijay-Otto}.
%existence and violation of such bounds were also shown to exist . In contrast, for the refrigerator case violation of the lower bound $\eta^{(2)}_{\rm ref}$ was pointed out in Ref.~\cite{Bijay-Otto}.  We see an exactly similar trend in our case of periodically driven continuous engines}.\\

For the upper bound in the engine regime, we provide a scatter plot and observe the validity of the corresponding bound. In Fig. \ref{upper-engine}, we plot the difference between $\eta_C^2$ and $\eta_{\rm eng}^{(2)}$ with $\Delta$ for $10^5$ randomly generated set of parameters within the range $(0,10)$, satisfying the constraints $\beta_h<\beta_c$ and $\Delta<\Delta_{\rm cr}$, for each point. From the figure it is evident that the difference is always positive, implying the validity of the upper bound in the engine regime. For the refrigerator regime, we have already argued in Eq. (\ref{upper-bound-refrigerator}) that the upper bound is respected.
It saturates in the limit when the entropy production rate vanishes, as illustrated in Fig. \ref{upper-saturate}, the upper bound saturates for both engine and refrigerator case in the limit of zero power which corresponds to $\Delta\rightarrow \Delta_{\rm cr}$. We emphasize that the above results are specific to sinusoidal modulation. In Sec.~\ref{secCRAB} we show that in case of more generic modulation schemes,  the lower bound for engine, and the upper bound for refrigerator always remain robust.

%%%%%%%%%%%%%%%%%%%%%%%%%%%%%%%%%%%%%%%%%%%%%%%%%%%%%%%%%%%%%%%%%%%%%%%%%%%%%%%%%%%%%%%%%%%%%%%%%%%%%%%%%%%%%%%%%%

\subsection{TUR minimization through optimal control}
\label{secCRAB}
\subsubsection{\textbf{\textit{CRAB Optimization Protocol }}}
\label{secCRAB1}
The sinusoidal modulation discussed above does not guarantee optimal operation. We, therefore now focus on the CRAB optimization protocol  \cite{caneva2011chopped,doria2011optimal, muller_2022}, aimed at minimizing the TUR ratio for heat currents and power. The CRAB optimization scheme has proven to be highly successful in several theoretical \cite{mukherjee13speeding, caneva14complexity}, as well as experimental \cite{omran19generation, borselli21two} works, and can be modeled as a generic periodic
modulation of $\omega(t)$ expressed as a truncated Fourier series:
\ba
\omega(t) &=& \omega_{0}+\frac{\mu}{2 N R(t)} \sum_{n=1}^{N}\Big[a_{n} \cos \left(\frac{2 \pi n t}{T}\right) \non\\ &+& b_{n} \sin \left(\frac{2 \pi n t}{T}\right)\Big],
\label{CRABeqn-1}
\ea
where the Hamiltonian is given by Eq. \eqref{Eq:hamil}.
Here the positive integer $N$ denotes the total number of frequencies considered, while $T$ is the time period of modulation.  The function $R(t)\rightarrow \infty$ for $t = 0$ and $t =T$, while $R(t) = 1$ for intermediate times, such that $\omega(t)=\omega_0$ at the beginning and end of a cycle.  We numerically optimize
the Fourier coefficients $-1 \leq a_n\leq 1$,  $-1 \leq b_n \leq 1$,  so as to optimize a relevant cost function, subject to certain constraints, such as the strength of control $\mu$.
In the present case, we perform the above optimization protocol in order to minimize the TUR ratio $\mathcal{R}_j$, as defined in Eq. \eqref{eqTUR}, for various currents and power (see also Appendix \ref{appec}). For the optimization, we choose the same spectral function as done for the sinusoidal case [Eq.~(\ref{GLorenz})].
%%%%%%%%%%%%%%%%%%%%%%%%%%%%
\begin{figure*}
    \centering
    \includegraphics[width=\textwidth]{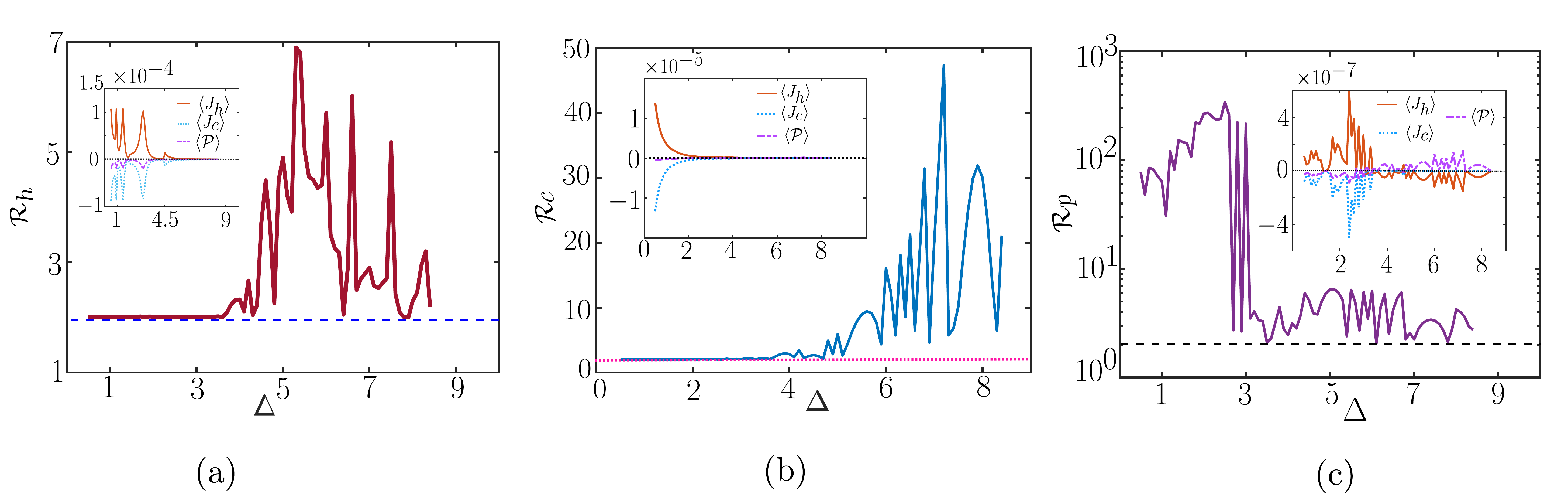}
    \caption{Plot  showing the TUR ratio (a) $\mathcal{R}_h$ for the hot current $J_h$, (b) $\mathcal{R}_c$ for the cold current $J_c$ and (c) $\mathcal{R}_{p}$ for the output power $\mathcal{P}$, as a function of $\Delta$, for the CRAB modulation Eq. \eqref{CRABeqn-1}. As expected, $\mathcal{R}_j$ is bounded from below by $\mathcal{R}_j = 2$  for all values of $\Delta$, for $j = h, c$, and $\mathcal{P}$.  The insets show  the variation of the mean energy currents $\langle J_h \rangle, \langle J_c \rangle$ and the mean power $\langle \mathcal{P} \rangle$ as  functions of $\Delta$.  As seen in (a) and (b), through CRAB optimization, one can have $\mathcal{R}_{h,c} \to 2^{+}$ at non-zero output power,  for a wide range of $\Delta$. %In contrast, numerical optimization suggests minimizing the TUR ratio diminishes the mean output power  as well. 
The parameters are fixed as $\omega_0=30$,  $\beta_h=0.005$, $\beta_c=0.01$, $\Gamma=0.2$, $\delta=3$, $\gamma_0=1$, $\mu=1$, and $N=10$.}
\label{Fig.4}
\end{figure*}
%%%%%%%%%%%%%%%%%%%%%%%%%%%%
\begin{figure*}
    \centering
    \includegraphics[width=\textwidth]{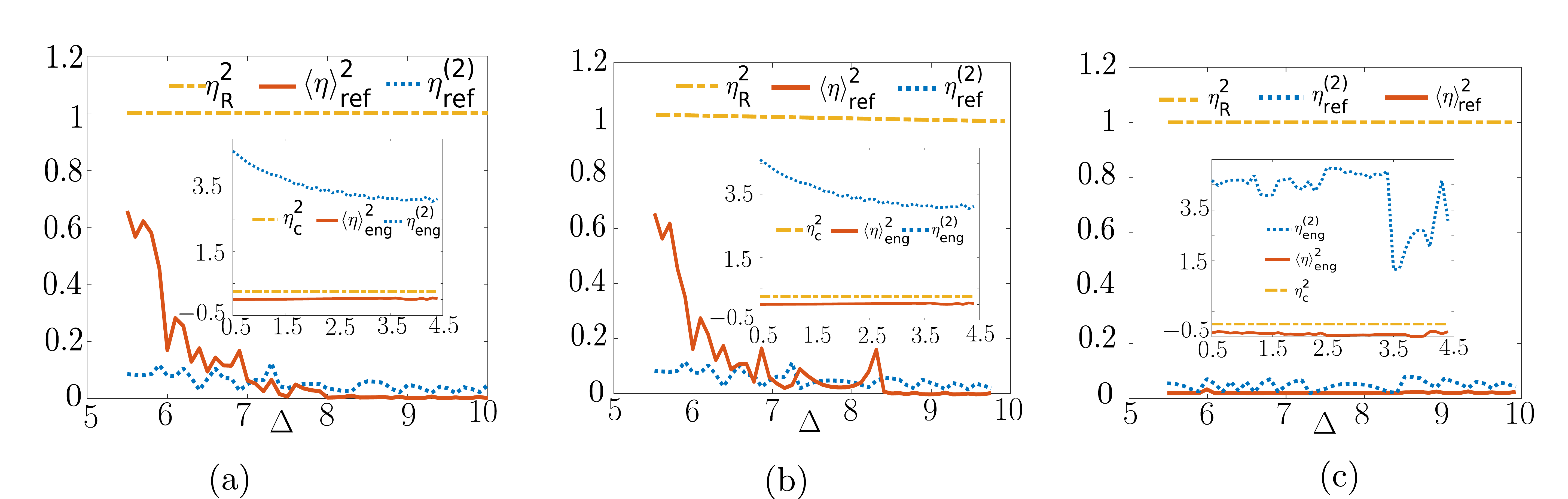}
    \caption{Plot of  $\langle\eta\rangle_{\rm ref}^2$,  $\eta_{\rm ref}^{(2)}$, and $\eta_R^{2}$   with $\Delta$, in the refrigerator regime. In Fig. \ref{Fig5}(a), for any $\Delta$, we have considered the pulse used in Fig. \ref{Fig.4}(a), aimed at minimizing $\mathcal{R}_h$ for that particular $\Delta$. Similarly, in Fig. \ref{Fig5}(b) and Fig. \ref{Fig5}(c) we have considered the pulse used in Fig. \ref{Fig.4}(b) and Fig. \ref{Fig.4}(c), respectively. The insets show the plots for of $\eta^{(2)}_{\rm eng},  \langle\eta\rangle_{\rm eng}^2$, and $\eta_C^{2} $ with $\Delta $ in the heat engine regime. We observe that in all cases the lower bound for the engine i.e., $\eta^{(2)}_{\rm eng} \ge \langle \eta \rangle^2$ and the upper bound for the refrigerator i.e., $\eta^{(2)}_{\rm ref} \le \eta_R^2$ always remain valid.}
    \label{Fig5}
\end{figure*}
%%%%%%%%%%%%%%%%%%%%%%%%%%%%%%%%%%%%%%%%%%%%%%%%%%%%%%%%%%%%%%%%%%%%%%%%%%%%%%%%%%%%%%%
\subsubsection{\textbf{\textit{Optimization and Result}}}
\label{secCRAB2}
For any modulation frequency $\Delta = 2\pi/T$, we numerically optimize the coefficients $\{a_n\},\{b_n\}$ through pattern search optimization method,  in order to  minimize the  TUR ratio for $J_h$ ($\mathcal{R}_h$, see Fig. \ref{Fig.4}(a)), $J_c$ ($\mathcal{R}_c$, see Fig. \ref{Fig.4}(b)) and $\mathcal{P}$ ($\mathcal{R}_{\mathcal{P}}$, see Fig. \ref{Fig.4}(c)) for that $\Delta$.
Here $n=1,2,3, ..., N$, and we consider a cutoff  $N=10$. The optimal set of coefficients $\{a_n, b_n\}_{\Delta}$ defines the periodic modulation with the corresponding  $\Delta$. 
As seen in Fig.~\ref{Fig.4}(a), Fig.~\ref{Fig.4}(b), and Fig.~\ref{Fig.4}(c) the TUR ratio $\mathcal{R}_j \geq 2$ for all $\Delta$, for $j = h, c$, and $\mathcal{P}$. Furthermore, we note that in the case of the sinusoidal modulation, the TUR ratio was minimal at the crossover point where all the currents were zero (see Fig. \ref{upper-saturate}). However the numerical optimization scheme suggests one can use the CRAB protocol to get a small TUR ratio ($\mathcal{R}_j \to 2^{+}$) for $J_h$ and $J_c$  at non-zero output power for a wide range of modulation frequencies $\Delta$ in the heat engine regime ($\langle J_h \rangle > 0, \langle J_c \rangle < 0, \langle \mathcal{P} \rangle < 0$), thus highlighting the advantage of performing CRAB optimization.
Moreover, a noteworthy observation in Fig. \ref{Fig.4}(a) is that  in the heat engine regime $(0< \Delta \lesssim 4)$, the TUR ratio for the input  i.e., $\mathcal{R}_h$ can be reduced to small values and can saturate the TUR bound through optimal
control, whereas the same may not be always true in case of the output heat current $\mathcal{J}_c$ in the refrigerator regime $\Delta \gtrsim 4$, as shown by  $\mathcal{R}_c$ in  Fig. \ref{Fig.4}(c).  This may be attributed to small values of $\langle \mathcal{J}_c \rangle$ in the refrigerator regime (cf. inset of Fig. \ref{Fig.4}(b)), as compared to larger values of $\langle \mathcal{J}_h \rangle$ in the heat engine regime (cf. inset of Fig. \ref{Fig.4}(a)). A similar behavior is also noticed for the TUR ratio $\mathcal{R}_p$ as well. Furthermore, the contrasting behaviors of TUR in the heat engine ($0 < \Delta \lesssim 4$) and the refrigerator ($\Delta \gtrsim 4$) regimes in Figs. \ref{Fig.4}(a), \ref{Fig.4}(b) and \ref{Fig.4}(c) suggest that TUR can exhibit signatures of heat engine to refrigerator transition for more generic modulation schemes.
We next study the bounds in the fluctuations in efficiencies in Figs. \ref{Fig5}(a)-\ref{Fig5}(c), for the pulses used in Figs. \ref{Fig.4}(a)- \ref{Fig.4}(c).
%. We arrive at similar conclusions for the bounds following Fig. \ref{fig:fig-02} and Fig. \ref{fig:fig-03}. 
As seen in Fig. \ref{Fig5}(a), the lower bound in the refrigerator regime, as suggested by Eq.~\eqref{eq:effref}, is violated for a wide range of $\Delta$,  while the upper bound i.e., $\eta^{(2)}_{\rm ref} \le \eta_R^2$ always remains  valid. This is in agreement with our  observation for the case of sinusoidal modulation (see Eq. \eqref{upper-bound-refrigerator}). Interestingly, in the engine regime, a completely opposite trend is observed for the bounds. We observe that the lower bound i.e., $\eta^{(2)}_{\rm eng} \ge \langle \eta \rangle^2$  always remains valid, as was also the case for sinusoidal driving (see Eq.~\eqref{eng-low-bound}); in contrast, now the upper bound gets violated for this optimal driving scenario. In summary, in all the cases we observe the validity of the lower bound for the engine  and the upper bound for the refrigerator. This is in stark contrast to autonomous quantum thermal machines, where the upper and lower bounds for fluctuations have been shown to hold in the linear response regime \cite{Bijay-machine-ss,Bijay-Otto}. On the other hand, the absence of linear response and the presence of external periodic modulation invalidates those bounds in the present context. These results indicate crucial differences in the nature of the fluctuations for currents in the engine and the refrigerator regimes, as well as in non-autonomous continuous thermal machines in comparison to their autonomous counterparts. A comparative analysis of fluctuations and optimization in continuous and stroke thermal machines may yield interesting results as well \cite{holubec-cycling}. Furthermore, they emphasize the importance of rigorous studies regarding bounds of fluctuations in generic quantum thermal machines. 
%%%%%%%%%%%%%%%%%%%%%%%%%%%%%%%%%%%%%%%%%%%%%%%%%%%%%%%%%%%%%%%%%%%%%%%%%%%%%%%%%
\subsection{Circular modulation} 
\label{section3D}
We next consider circular modulation, given by:
\begin{equation}
\label{driven-H}
H_S(t)=\frac{\omega_0}{2}\sigma_z+g(\sigma^-e^{i\Omega t}+\sigma^+ e^{-i\Omega t}).
\end{equation}
 In contrast to the sinusoidal and the CRAB modulations discussed above, here $\left[H_{\rm S}(t), H_{\rm S}(t^{\prime}) \right]$ may not commute for $t \neq t^{\prime}$. This non-commutative property of the system Hamiltonian can lead to quantum friction, thereby changing the qualitative nature of the thermal machine \cite{kosloff02discrete}. Consequently, such modulation \eqref{driven-H} can play an important role in understanding the role of fluctuations in continuous minimal quantum thermal machines. Moreover, as we show below, the model considered here results in the realization of a heat accelerator, thereby allowing us to extend our analysis to periodically modulated continuous thermal machines operating beyond the heat engine or refrigerator regimes \cite{mukherjee16speed}.

One can refer to the Floquet analysis in Appendix~\ref{appflq} and Appendix~\ref{appc} to arrive at the following generalized master equation \cite{secular-2,szczygielski13}, starting from Eq. (\ref{Gen-master-eq}),
\begin{widetext} 
\begin{align}
\nonumber
 \dot{\tilde{\rho}}_{\alpha\beta}=&-\sum_{j=h,c}\int_0^{\infty} d\tau \Phi_j(\tau)\sum_{m,n,l,l'}e^{i(\epsilon_\alpha-\epsilon_m)t}e^{i(\epsilon_m-\epsilon_n)(t-\tau)}e^{il\Omega t}e^{il'\Omega (t-\tau)}S_{\alpha m}^l S_{mn}^{l'}\rho_{n\beta}\\
\nonumber
& +\sum_{j=h,c}\int_0^{\infty} d\tau \Phi_j(-2\chi,-\tau)\sum_{m,n,l,l'}e^{i(\epsilon_\alpha-\epsilon_m)t}e^{i(\epsilon_n-\epsilon_\beta)(t-\tau)}e^{il\Omega t}e^{il'\Omega (t-\tau)}S_{\alpha m}^l S_{n\beta}^{l'}\rho_{mn}\\
\nonumber
& +\sum_{j=h,c}\int_0^{\infty} d\tau \Phi_j(-2\chi,\tau)\sum_{m,n,l,l'}e^{i(\epsilon_n-\epsilon_\beta)t}e^{i(\epsilon_\alpha-\epsilon_m)(t-\tau)}e^{il\Omega (t-\tau)}e^{il'\Omega t}S_{\alpha m}^l S_{n \beta}^{l'}\rho_{mn}\\
&-\sum_{j=h,c}\int_0^{\infty} d\tau \Phi_j(-\tau)\sum_{m,n,l,l'}e^{i(\epsilon_m-\epsilon_n)(t-\tau)}e^{-i(\epsilon_n-\epsilon_\beta)t}e^{il\Omega (t-\tau)}e^{il'\Omega t}S_{mn}^l S_{n\beta}^{l'}\rho_{\alpha m}.
\end{align}
\end{widetext}
where, $\epsilon_\alpha$'s are the eigenvalues of the Floquet Hamiltonian $H_F$, and all the indices are integers (Appendix~\ref{appflq}).
We here perform the secular approximations for the above master equation; we neglect the fast oscillating terms of the form $e^{il\Omega t}$ (for $l\neq 0$). Additionally, we also neglect the terms of the form of $e^{i(\epsilon_\alpha-\epsilon_\beta)t}$ (for $\alpha \neq \beta$) \cite{secular-1, secular-2}.   As a result of the secular approximation, evolution of diagonal and off-diagonal terms of the above master equation get decoupled, and we get a conventional Pauli rate equation with the diagonal entries containing time-independent transition rates. We therefore have,
\begin{equation}
\label{floquet-rate}
\begin{pmatrix}
\dot{\tilde{\rho}}_{00}(\chi,t)\\
\dot{\tilde{\rho}}_{11}(\chi,t)
\end{pmatrix}
=
\begin{pmatrix}
l_{00}^{\chi} & l_{01}^{\chi}\\
l_{10}^{\chi} & l_{11}^\chi
\end{pmatrix}
\begin{pmatrix}
\tilde{\rho}_{00}(\chi,t)\\
\tilde{\rho}_{11}(\chi,t)
\end{pmatrix},
\end{equation}
where,
\begin{align}
\nonumber
l_{00}^{\chi}=&{|S_{11}^1|}^2\sum_{j=h,c}\left[G_j(\Omega)(e^{-i\Omega \chi_j}-1)+G_j(-\Omega)(e^{i\Omega \chi_j}-1)\right]\\
&-{|S_{12}^1|}^2 \sum_{j=h,c} G_j(\Omega\!-\!\Omega_R)\!-\!{|S_{21}^1|}^2  \sum_{j=h,c} G_j(-\Omega\!-\!\Omega_R)
\end{align}
\begin{align}
\nonumber
l_{01}^{\chi}=&{|S_{12}^1|}^2\sum_{j=h,c}G_j(-\Omega+\Omega_R)e^{-i(-\Omega+\Omega_R)\chi_j}\\
&+{|S_{21}^1|}^2\sum_{j=h,c}G_j(\Omega+\Omega_R)e^{-i(\Omega+\Omega_R)\chi_j}
\end{align}
\begin{align}
\nonumber
l_{10}^{\chi}=&{|S_{12}^1|}^2\sum_{j=h,c}G_j(\Omega-\Omega_R)e^{-i(\Omega-\Omega_R)\chi_j}\\
&+{|S_{21}^1|}^2\sum_{j=h,c}G_j(-\Omega-\Omega_R)e^{i(\Omega+\Omega_R)\chi_j}
\end{align}
\begin{align}
\nonumber
l_{11}^{\chi}=&{|S_{22}^1|}^2\sum_{j=h,c}\left[G_j(\Omega)(e^{-i\Omega \chi_j}-1)+G_j(-\Omega)(e^{i\Omega \chi_j}-1)\right]\\
&-{|S_{12}^1|}^2 \sum_{j=h,c} G_j(-\Omega+\Omega_R)-{|S_{21}^1|}^2\, \sum_{j=h,c} G_j(\Omega+\Omega_R)
\end{align}
As before, we are interested in the long time limit and computing the steady-state currents and fluctuations. We once again use the properties of the cumulant generating function (Eq. (\ref{cumulantss}), Eq. (\ref{formula1}) and Eq. (\ref{formula2})) to calculate the mean and variance of the heat currents and the power and check the validity of the inequality \eqref{eqTURineq}.  
\begin{figure}
\begin{center}
\includegraphics[width=\columnwidth]{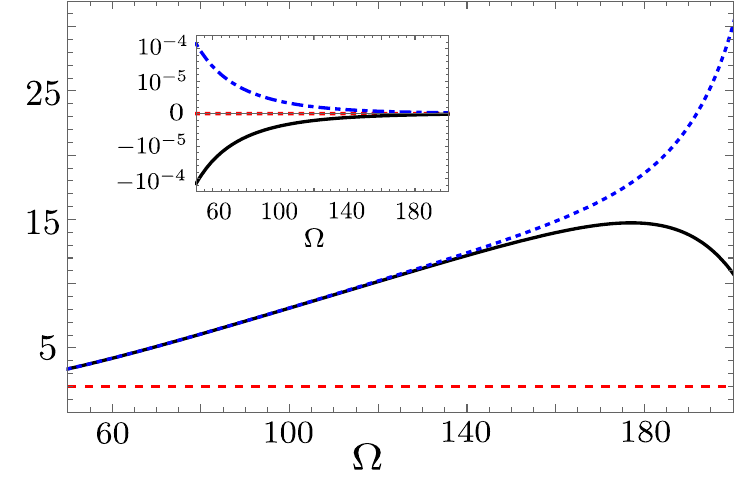}
\caption{Plot of TUR ratio for hot (black solid) and cold (blue dotted line) bath currents as a function of driving frequency $\Omega$, for circular modulation. Here, $\omega_0=25$, $\beta_h=0.01$, $\beta_c=0.06$, $g=0.02$, $\Gamma=0.2$, $\delta=3$, and $\gamma_0=1$.  The dashed horizontal red line represents the value of the lower bound (2) of TUR ratio. The inset shows the hot (blue dashed-dotted), cold (black solid line) bath currents, and the power (red dotted line).}
\label{currents-driv}
\end{center}
\end{figure}
Here we consider a Lorentzian bath spectrum for both the hot bath and  the cold bath as follows
\begin{align}
&G_j(\omega)=\frac{\gamma_0\Gamma^2}{{(\delta-\omega)}^2+\Gamma^2},\\
& G_j(-\omega)=G_j(\omega)e^{-\beta_j\omega}, ~~~j=\{h,c\}
\end{align}
where the parameters are defined after Eq. \eqref{GLorenz}; $G_h(\omega)$ and $G_c(\omega)$ assume maxima at $\omega = \delta.$
%Here, $j=\{h,c\}$. 
As shown in the inset of Fig. \ref{currents-driv}, the machine always works as a heat accelerator for all values of $\Omega$. Thermal machines working on time scales in which the typical Born-Markov approximations do
not hold may lead to different possibilities \cite{secular-3, arpan}.  As we don't get any useful work from this machine, we are not interested in the fluctuations of power and all the precision bounds introduced in the previous section. Instead, we solely focus on the TUR ratio for hot and cold heat currents.
As shown in Fig.~\ref{currents-driv}, the inequality \eqref{eqTURineq} is satisfied in this scenario. This can be expected as due to the secular approximation, the diagonal and off-diagonal elements of the system density matrix evolve independently in time which results in a Pauli-type master equation with time-independent coefficients for the occupation probabilities. This, along with the local detailed balance condition, results in the condition $\mathcal{R} \geq 2$ \cite{tur-satis2, Proesmans_2019,tur-satis3, tur-satis}.
Unlike the case of sinusoidal modulation, the noise to signal ratio is not the same for hot and cold currents in case of circular modulation, as can be seen from Fig.~\ref{currents-driv}. For sinusoidal modulation, the spectral separation condition Eq. (\ref{eqG}) results in the difference between noise to signal ratio for the hot and cold current to vanish, as shown in Appendix \ref{appena}. However, the same is not true in case of circular modulation.

%%%%%%%%%%%%%%%%%%%%%%%%%%%%

\section{Conclusion}
\label{conclu}
In this paper, we discuss periodically driven continuous heat machines from the Floquet perspective.  Employing the counting field statistics approach, we compute the steady-state heat currents and the associated fluctuations for generic periodically driven continuous thermal machines with the two-level systems as a working medium. We have exemplified our theory using the specific cases of sinusoidal, CRAB optimized and circular modulations. We have analyzed different precision bounds and trade-offs with these fluctuations; specifically, we find that TUR is satisfied for the machines considered here.
One can operate the sinusoidally modulated thermal machine as a heat engine, or a refrigerator, depending on the frequency of modulation, thus motivating us to study the fluctuations in this model in greater detail. Interestingly, our analysis shows that the noise-to-signal ratio for the heat currents from the hot and the cold baths are equal in  case of sinusoidal modulation. 
Moreover, one can define a parameter to quantify the fluctuations in efficiency, which appears to be bounded from both above and below in the heat engine regime, while the existence of an equivalent lower bound is not clear in the case of the refrigerator regime. Note that, similar phenomena was observed recently for a discrete stroke heat machine as well \cite{Bijay-Otto}. 

We have used CRAB optimization protocol to minimize TUR and study the bounds in the fluctuations of efficiency. As expected, the TUR is always bounded from below by two. Numerical analysis suggests that (i)  in the heat engine regime, one can use optimal control to operate the thermal machine with minimum TUR ($= 2$) for the heat currents, at non-zero power output; (ii) one can reduce the TUR ratio for the input (heat current absorbed from the hot bath in the engine regime and power in the refrigerator
regime)  to small values through optimal
control, whereas the same may not be always true in case
of the output (power output in the engine regime and
heat current extracted from the cold bath in the refrigerator regime) (see Fig. \ref{Fig.4}); (iii) the lower (upper) bound for the fluctuations in efficiency is always satisfied in the heat engine (refrigerator) regime (see Eq. \eqref{eq:effref} and Fig. \ref{Fig5}).

Finally, we have used circular modulation to show  that the TUR ratio is satisfied in the case of a continuously driven heat accelerator as well. 

The results presented in this work highlight the importance of optimal control to study fundamental bounds in the performance of quantum machines, and for designing high-performing quantum devices.
We note that violation of the TUR bound Eq. \eqref{eqTURineq}  has been reported in similar models described by phenomenological master equation and in presence of coherent dynamics \cite{coherent-4, TURDvira, tur-satis, coherent-2, coherent-3, coherent-5}. Consequently, the validity of the TUR bound in absence of secular approximation is an intriguing open question, which we plan to address in future works. The validity/existence of the bounds for fluctuations in broad classes of quantum machines is also worth looking at.
\\\\
%%%%%%%%%%%%%%%%%%%%%%%%%%%%%%%%%%%%%%%%%%%%%%%%%%%%%%%%%%%%%%%%%%%%%%%%%%%%%%%%%%%%%%%%%%%%%%%%%%%%%%%%%%%%%%%%%%
\section*{ACKNOWLEDGMENTS}
AD acknowledges the support of Post Doctoral Fellowship at NCU, Toru\'n, Poland. BKA acknowledges the MATRICS grant MTR/2020/000472 from SERB, Government of India and the Shastri Indo-Canadian Institute for providing financial support for this research work in the form of a Shastri Institutional Collaborative Research Grant (SICRG). V.M. acknowledges support from Science and Engineering Research Board (SERB) through MATRICS (Project No.
MTR/2021/000055) and Seed Grant from IISER Berhampur. SM and VM acknowledge
funding support for Chanakya - PG fellowship (Project No. I-HUB/PGF/2021-22/019) from the National Mission on Interdisciplinary Cyber Physical Systems, of the Department of Science and Technology, Govt. of India, through the I-HUB Quantum Technology
Foundation. 
\onecolumngrid
\appendix

\section{Floquet theory}
\label{appflq}{
In this section we assume a periodically modulated Hamiltonian $H_S(t)$ with time period $T$, i.e., $H_S(t)=H_S(t+T)$ where, $T=\frac{2\pi}{\Omega}$. Solution of Schr{\"o}dinger equation with this Hamiltonian reads as,
\begin{equation}
\ket{\psi(t)}=U(t,t_0)\ket{\psi(t_0)},
\end{equation}
where $U(t,t_0)= \mathcal{T}e^{-i\int_{t_0}^t  H(\tau)d\tau}$. Owing to the periodicity of the Hamiltonian $H(t)$, one can show that \cite{floquet-1, floquet-2,floquet-3,floquet-4},
\begin{equation}
 U(t_0+nT,t_0)={[U(t_0+T,t_0)]}^n
\end{equation}
The unitary Floquet propagator $U(t_0+T,t_0)$ can be written as
\begin{equation}
U(t_0+T,t_0)=e^{-iH_F[t_0]T},
\end{equation}
where $H_F$ is called the Floquet Hamiltonian \cite{klimovsky15thermodynamics}. Clearly, $H_F[t_0]$ depends on our choice of $t_0$. Consequently, to avoid confusion we choose $t_0 = 0$, and write $U(T)=e^{-iH_F T}$.

 The next quantity we are interested in is the unitary propagator for arbitrary time $t$. One can show that,
\begin{equation}
U(t,0)\equiv U(t)=P(t)e^{-iH_F t}, ~~ \text{where}, ~~ P(t+T)=P(t) = U(t)e^{i H_F t}.
\end{equation}
$P(t)$ is called the kick operator. Denoting the complete set of eigenvectors for $U(T)$ by $\{\ket{\phi_j}\}$, one can write
\begin{equation}
U(T)\ket{\phi_j}=e^{-i\epsilon_j T}\ket{\phi_j}\Rightarrow H_F=\sum_j \epsilon_j\ket{\phi_j}\bra{\phi_j}.
\end{equation}
Further,
\begin{equation}
U(t)\ket{\phi_j}=P(t)e^{-iH_F t}\ket{\phi_j}=e^{-i\epsilon_j t}\ket{\phi_j(t)},
\end{equation}
where, $\ket{\phi_j(t)}=P(t)\ket{\phi_j}$, with $\ket{\phi_j(t+T)}=\ket{\phi_j(t)}$. Any initial state can be written as the linear combination of the Floquet modes $\ket{\phi_j}$,
\begin{equation}
\ket{\psi(t_0)}=\sum_j c_j\ket{\phi_j},
\end{equation}
Clearly, $\ket{\psi(t)}=\sum_j c_j e^{-i\epsilon_j t}\ket{\phi_j(t)}$ is a solution of Schr{\"o}dinger equation. One can express $U(t)$ and $P(t)$ as,
\begin{equation}
U(t)=\sum_j e^{-i\epsilon_j t}\ket{\phi_j(t)}\bra{\phi_j},~~ P(t)=\sum_j \ket{\phi_j(t)}\bra{\phi_j}.
\end{equation}
To evaluate the term $\tilde{S}(t)=U^{\dagger}_S(t)S U_S(t)$ in Eq. (\ref{parent-eq}), we proceed as follows,
\begin{align}
\nonumber
\tilde{S}(t)&=\sum_{j,k}e^{i(\epsilon_k-\epsilon_j)t}\bra{\phi_k(t)}S\ket{\phi_j(t)}\left(\ket{\phi_k}\bra{\phi_j}\right),\\
\label{s-expres}
&=\sum_{j,k,\alpha}e^{i(\epsilon_k-\epsilon_j)t}e^{i\alpha\Omega t}S_{kj}^{\alpha}\left(\ket{\phi_k}\bra{\phi_j}\right)
\end{align}
where, 
\begin{equation}
S_{kj}^\alpha=\left[\frac{1}{T}\int_0^T dt \bra{\phi_k(t)}S\ket{\phi_j(t)}e^{-i\alpha\Omega t}\right],
\end{equation}
are the Fourier components of the periodic function $\bra{\phi_k(t)}S\ket{\phi_j(t)}$. Now, for the Hamiltonian in Eq. (\ref{model-hamil}), $H_F=\frac{1}{2}\omega_0\sigma_z$, and $P(t)=\frac{1}{2}\lambda\Delta\sin(\Delta t)\sigma_z$. This implies $\ket{\phi_1}$ and $\ket{\phi_2}$ are $\ket{0}$ and $\ket{1}$ respectively, where, $\sigma_z\ket{0}=-\ket{0}$, and $\sigma_z\ket{1}=\ket{1}$. Hence,
\begin{align}
e^{-iH_F t}=e^{i\omega_0 t/2}\ket{0}\bra{0}+e^{-i\omega_0 t/2}\ket{1}\bra{1}.
\end{align}
Using this, we can write,
\begin{align}
\nonumber
\tilde{\sigma}_x(t)=e^{-i\omega_0 t}&\bra{0}P^\dagger(t)\sigma_x P(t)\ket{1}(\ket{0}\bra{1})\\
& +e^{i\omega_0 t}\bra{1} P^\dagger(t)\sigma_x P(t)\ket{0}(\ket{1}\bra{0})
\end{align}
Putting $P(t)=\sum_j \ket{\phi_j(t)}\bra{\phi_j}$, one can see that this equation is exactly same as Eq. (\ref{s-expres}). Now, from the Fourier components of the terms $\bra{i} P^\dagger(t)\sigma_x P(t)\ket{j}$ and noting that $\ket{0}\bra{1}=\sigma^-=\frac{1}{2}(\sigma_x-i\sigma_y)$ and $\ket{1}\bra{0}=\sigma^+=\frac{1}{2} (\sigma_x+i\sigma_y)$, we recover the Eq. (\ref{x-expr}), where,
\begin{equation}
\eta(q)=\left[\frac{1}{T}\int_0^T dt \bra{0}P^\dagger(t)\sigma_x P(t)\ket{1}e^{-iq\Delta t}\right],
\end{equation}
are the Fourier components.

\section{Expressions for mean currents and fluctuations under sinusoidal modulation}
\label{appena}
In this section, we provide expressions for the mean currents and the associated fluctuations for the sinusoidal driving case. From Eq. (\ref{formula1}), we get,
\ba
\langle J_h\rangle&=&-\frac{\lambda^2}{4}\frac{(\omega_0+\Delta)G_h(\omega_0+\Delta)G_c(\omega_0-\Delta)\Big(e^{-\beta_c(\omega_0-\Delta)}-e^{-\beta_h(\omega_0+\Delta)}\Big)}{G_h(\omega_0+\Delta)\Big(1+e^{-\beta_h(\omega_0+\Delta)}\Big)+G_c(\omega_0-\Delta)\Big(1+e^{-\beta_c(\omega_0-\Delta)}\Big)},\non\\
\langle J_c\rangle&=&-\frac{\lambda^2}{4}\frac{(\omega_0-\Delta)G_h(\omega_0+\Delta)G_c(\omega_0-\Delta)\Big(e^{-\beta_c(\omega_0-\Delta)}-e^{-\beta_h(\omega_0+\Delta)}\Big)}{G_h(\omega_0+\Delta)\Big(1+e^{-\beta_h(\omega_0+\Delta)}\Big)+G_c(\omega_0-\Delta)\Big(1+e^{-\beta_c(\omega_0-\Delta)}\Big)}.
\ea
It can be easily seen from the above expressions that at $\Delta = \Delta_{cr}$ both these currents and as a result the power also vanishes \cite{klimovsky13minimal}.
Next, we provide expressions for the variance starting from Eq. (\ref{formula2}), given by,
%\begin{align}
%\nonumber
\ba
{\rm var}(J_h)&=
\frac{\lambda^2}{4}\frac{(\omega_0+\Delta)^2 G_h(\omega_0+\Delta)\Big[2e^{-\beta_h(\omega_0+\Delta)}G_h(\omega_0+\Delta)+G_c(\omega_0-\Delta)\Big(e^{-\beta_c(\omega_0-\Delta)}+e^{-\beta_h(\omega_0+\Delta)}\Big)\Big]}{G_h(\omega_0+\Delta)\Big(1+e^{-\beta_h(\omega_0+\Delta)}\Big)+G_c(\omega_0-\Delta)\Big(1+e^{-\beta_c(\omega_0-\Delta)}\Big)}\non\\
&-\frac{\lambda^2}{2}\Bigg[\frac{e^{-\beta_h(\omega_0+\Delta)}{(\omega_0+\Delta)}^2 G_h^2(\omega_0+\Delta)+16/\lambda^4\langle J_h\rangle^2}{G_h(\omega_0+\Delta)\big(1+e^{-\beta_h(\omega_0+\Delta)}\big)+G_c(\omega_0-\Delta)\Big(1+e^{-\beta_c(\omega_0-\Delta)}\Big)}\Bigg],\non\\
%\end{align}
%\begin{align}
%\nonumber
{\rm var}(J_c)&=
\frac{\lambda^2}{4}\frac{(\omega_0-\Delta)^2 G_c(\omega_0-\Delta)\Big[2e^{-\beta_c(\omega_0-\Delta)}G_c(\omega_0-\Delta)+G_h(\omega_0+\Delta)\Big(e^{-\beta_c(\omega_0-\Delta)}+e^{-\beta_h(\omega_0+\Delta)}\Big)\Big]}{G_h(\omega_0+\Delta)\Big(1+e^{-\beta_h(\omega_0+\Delta)}\Big)+G_c(\omega_0-\Delta)\Big(1+e^{-\beta_c(\omega_0-\Delta)}\Big)}\non\\
&-\frac{\lambda^2}{2}\Bigg[\frac{e^{-\beta_c(\omega_0-\Delta)}{(\omega_0-\Delta)}^2 G_c^2(\omega_0-\Delta)+16/\lambda^4\langle J_c\rangle^2}{G_h(\omega_0+\Delta)\big(1+e^{-\beta_h(\omega_0+\Delta)}\big)+G_c(\omega_0-\Delta)\Big(1+e^{-\beta_c(\omega_0-\Delta)}\Big)}\Bigg].
\ea
%\end{align}
Covariance term is given by $\langle J_h J_c\rangle-\langle J_h\rangle \langle J_c\rangle$, where,
\begin{align}
%\nonumber
\langle J_h J_c\rangle=-\frac{\lambda^2}{4}\Bigg[\frac{{(\omega_0+\Delta)}(\omega_0-\Delta) G_h(\omega_0+\Delta)G_c(\omega_0-\Delta)\big(e^{-\beta_h(\omega_0+\Delta)}+e^{-\beta_c(\omega_0-\Delta)}\big)+32/\lambda^4\langle J_h\rangle \langle J_c\rangle}{G_h(\omega_0+\Delta)\big(1+e^{-\beta_h(\omega_0+\Delta)}\big)+G_c(\omega_0-\Delta)\Big(1+e^{-\beta_c(\omega_0-\Delta)}\Big)}\Bigg]
\end{align}

\section{Mean current and fluctuations for CRAB modulation }
\label{appec}
In this section, we focus on the mean currents and fluctuations for the generic  periodic modulation used for CRAB optimization in Sec. \ref{secCRAB} (see the supplementary of  \cite{mukherjee19enhanced}). We note that in contrast to sinusoidal modulation (see Eqs. \eqref{eqsin} and \eqref{eqP}), in case of CRAB modulation one can have multiple Floquet modes with significant weights $P_q$. This can in turn result in negative $\omega_q=\omega_0+q\Delta$ for large $|q|$.
We can rewrite Eq.\eqref{formula1} as follows, 
\begin{align}
\label{formula4}
& \langle J_j\rangle= \left.\lim_{t\rightarrow \infty}\frac{d}{dt}\frac{\partial}{\partial(i\chi_j)}\mathcal{C}(\chi,t)\right\vert_{\chi=0}=\left.\frac{\partial \lambda(\chi)}{\partial(i\chi_j)}\right\vert_{\chi=0,}\non\\
&={\sum_q}^{+} \frac{P_q (\omega_0+q\Delta)}{w^{\prime}+1}G_j(\omega_0+q\Delta)\Big[e^{-\beta_j(\omega_0+q\Delta)}-w^{\prime}\Big]\non\\&+ {\sum_q}^{-} \frac{P_q (|\omega_0+q\Delta)|}{w^{\prime}+1}G_j(|\omega_0+q\Delta|)e^{-\beta_{j}(|\omega_0+q\Delta|)}\Big[e^{-\beta_j|\omega_0+q\Delta)|}-w^{\prime}\Big].
\end{align}
Here, ${\sum_{q}}^{+}$(${\sum_{q}}^{-})$ denotes summation over integer q such that $\omega_q\geq 0 (\omega_q < 0)$, $\omega_q=\omega_0+q\Delta$.
while, $w^{\prime}$ is the ratio of the diagonal entries of $\tilde{\rho}_S (0, t)$ in the steady state, and can be obtained straightforwardly as 
\begin{equation}
\label{p11p22ss}
\frac{p_1^{ss}}{p_2^{ss}}\equiv w^{\prime}=\frac{\sum_{q,j}^{+} P_q G_j(\omega_0+q\Delta)e^{-\beta_j(\omega_0+q\Delta)}+ \sum_{q,j}^{-} P_q G_j(|\omega_0+q\Delta|)}{\sum_{q,j}^{+}P_qG_j(\omega_0+q\Delta)+\sum_{q,j}^{-} P_qG_j(|\omega_0+q\Delta|)e^{-\beta_j(|\omega_0+q\Delta|)}}=\frac{l_{11}}{l_{00}}.
\end{equation}
To arrive at this expression we have used the KMS boundary condition,
\begin{equation}
G_j(-\omega_0-q\Delta)=e^{-\beta_j(\omega_0+q\Delta)}G_j(\omega_0+q\Delta) ; \quad \text{for} \,\,  \omega_q \geq 0.
\end{equation}
Similarly, the variation is given as,
\begin{align}
\label{formula21}
& {\rm var}(J_j)=\left.\lim_{t\rightarrow \infty}\frac{d}{dt}\frac{\partial^2}{\partial{(i\chi_j)}^2}\mathcal{C}(\chi,t)\right\vert_{\chi=0}=\left.\frac{\partial^2\lambda(\chi)}{\partial(i\chi_j)^2}\right\vert_{\chi=0}, \nonumber \\
\end{align}
which upon simplification leads to the following expression:
\begin{align}
{\rm var}(J_j)&={\sum_q}^{+} \frac{P_q {(\omega_0+q\Delta)}^2}{w^{\prime}+1}G_j(\omega_0+q\Delta)\Big[e^{-\beta_j(\omega_0+q\Delta)}+w^{\prime}\Big]\non\\
&+{\sum_q}^{-} \frac{P_q {(\omega_0+q\Delta)}^2}{w^{\prime}+1}G_j(|\omega_0+q\Delta|)e^{-\beta_j(|\omega_0+q\Delta|)}\Big[e^{-\beta_j(\omega_0+q\Delta)}+w^{\prime}\Big]\non\\
&-\frac{2{\langle J_j\rangle}^2}{\sum_{q,j}^{+}P_qG_j(\omega_0+q\Delta)\Big[e^{-\beta_j(\omega_0+q\Delta)}+1\Big]+\sum_{q,j}^{-}P_qG_j(|\omega_0+q\Delta|)e^{-\beta_j(|\omega_0+q\Delta|)}\Big[e^{-\beta_j(\omega_0+q\Delta)}+1\Big]}\non\\    
&-\frac{2}{\sum_{q,j}^{+}P_qG_j(\omega_0+q\Delta)\Big[e^{-\beta_j(\omega_0+q\Delta)}+1\Big]+\sum_{q,j}^{-}P_qG_j(|\omega_0+q\Delta|)e^{-\beta_j(|\omega_0+q\Delta|)}\Big[e^{-\beta_j(|\omega_0+q\Delta|)}+1\Big]}\non\\
&\times\Bigg[{\sum_{q',q''}}^{++}P_{q'} P_{q''}e^{-\beta_j(\omega_0+q'\Delta)}(\omega_0+q'\Delta)(\omega_0+q''\Delta)G_j(\omega_0+q'\Delta)G_j(\omega_0+q''\Delta)\non\\
&+ {\sum_{q',q''}}^{--}P_{q'} P_{q''}(\omega_0+q'\Delta)(\omega_0+q''\Delta)G_j(|\omega_0+q'\Delta|)G_j(|\omega_0+q''\Delta|) e^{-\beta_j(|\omega_0+q''\Delta|)}\non\\
&+ {\sum_{q',q''}}^{+-}P_{q'} P_{q''}(\omega_0+q'\Delta)(\omega_0+q''\Delta)G_j(\omega_0+q'\Delta)G_j(|\omega_0+q''\Delta|)e^{-\beta_j(|\omega_0(q'+q'')\Delta|)}\non\\
&+{\sum_{q',q''}}^{-+}P_{q'} P_{q''}(\omega_0+q'\Delta)(\omega_0+q''\Delta)G_j(|\omega_0+q'\Delta|)G_j(\omega_0+q''\Delta) \Bigg]. 
\end{align}
Where $\sum_{q',q''}^{++}$($ \sum_{q',q''}^{--}$) denotes summation over integer $q'$ and $q''$ such that $\omega_{q'}>0,\omega_{q''}>0$($\omega_{q'}<0,\omega_{q''}<0$) and $\sum_{q',q''}^{+-}$($ \sum_{q',q''}^{-+}$) denotes summation over the integer $q'$ and $q''$ such that $\omega_{q'}>0,\omega_{q''}<0$($\omega_{q'}<0,\omega_{q''}>0)$.
The covariance term is given by ${\rm Cov}(J_h,J_c)=\langle J_hJ_c\rangle-\langle J_h\rangle \langle J_c\rangle$, where,
\begin{align}
\langle J_hJ_c\rangle&=
\left.\frac{\partial^2 \lambda(\chi)}{\partial (i\chi_c)\partial(i\chi_h)}\right\vert_{\chi=0}
=\frac{1}{l_{00}+l_{11}}\Big(\left.\frac{\partial l_{10}^{\chi}}{\partial(i\chi_c)}\frac{\partial l_{01}^{\chi}}{\partial(i\chi_h)}+\frac{\partial l_{01}^{\chi}}{\partial(i\chi_c)}\frac{\partial l_{10}^{\chi}}{\partial(i\chi_h)}\Big)\right\vert_{\chi_j=0}-\frac{2}{l_{00}+l_{11}}\langle J_h\rangle \langle J_c \rangle;\non\\
&\Big(\left.\frac{\partial l_{10}^{\chi}}{\partial(i\chi_c)}\frac{\partial l_{01}^{\chi}}{\partial(i\chi_h)}+\frac{\partial l_{01}^{\chi}}{\partial(i\chi_c)}\frac{\partial l_{10}^{\chi}}{\partial(i\chi_h)}\Big)\right\vert_{\chi_j=0}=\non\\ 
&-{\sum_{q',q''}}^{++} P_{q'}  P_{q''}e^{-\beta_h(\omega_0+q'\Delta)}(\omega_0+q'\Delta)(\omega_0+q''\Delta)G_h(\omega_0+q'\Delta)G_c(\omega_0+q''\Delta)\non\\
&-{\sum_{q',q''}}^{++}  P_{q'} P_{q''}e^{-\beta_c(\omega_0+q''\Delta)}(\omega_0+q'\Delta)(\omega_0+q''\Delta)G_h(\omega_0+q'\Delta)G_c(\omega_0+q''\Delta)\non\\
&-{\sum_{q',q''}}^{--}  P_{q'}  P_{q''}e^{-\beta_c(|\omega_0+q''\Delta|)}(\omega_0+q'\Delta)(\omega_0+q''\Delta)G_h(\omega_0+q'\Delta)G_c(|\omega_0+q''\Delta|)\non\\
&-{\sum_{q',q''}}^{--} P_{q'}  P_{q''}e^{-\beta_h(|\omega_0+q'\Delta)|}(\omega_0+q'\Delta)(\omega_0+q''\Delta)G_h(|\omega_0+q'\Delta|)G_c(|\omega_0+q''\Delta|)\non \\
&- {\sum_{q',q''}}^{+-} P_{q'}  P_{q''}e^{-\beta_h(|\omega_0+q'\Delta)|}e^{-\beta_c(|\omega_0+q''\Delta)|}(\omega_0+q'\Delta)(\omega_0+q''\Delta)G_h(|\omega_0+q'\Delta|)G_c(|\omega_0+q''\Delta|)\non\\
&-{\sum_{q',q''}}^{+-} P_{q'}  P_{q''}(\omega_0+q'\Delta)(\omega_0+q''\Delta)G_h(|\omega_0+q'\Delta|)G_c(|\omega_0+q''\Delta|)\non\\
&- {\sum_{q',q''}}^{-+} P_{q'}  P_{q''}(\omega_0+q'\Delta)(\omega_0+q''\Delta)G_h(|\omega_0+q'\Delta|)G_c(|\omega_0+q''\Delta|)\non\\
&+{\sum_{q',q''}}^{-+} P_{q'}  P_{q''} e^{-\beta_h(|\omega_0+q'\Delta)|}e^{-\beta_c(|\omega_0+q''\Delta)|}(\omega_0+q'\Delta)(\omega_0+q''\Delta)G_h(|\omega_0+q'\Delta|)G_c(|\omega_0+q''\Delta|).
\end{align}

\section{Floquet analysis for circular modulation}
\label{appc}
Here we summarise the Floquet analysis for the circular driving case. In this case, the Floquet Hamiltonian and the kick operator are given as,
\begin{align}
& H_F=\left(\frac{\omega_0-\Omega}{2}\right)\sigma_z+g\sigma_x-\frac{\Omega}{2}\mathds{1},\\
& P(t)=e^{-\frac{i\Omega t}{2}\left(\sigma_z+\mathds{1}\right)}.
\end{align} 
Following the previous discussion, we calculate the Floquet eigenmodes and eigen-energies by the eigenvalue decomposition of $H_F$: $\sum_{j=1}^2 \epsilon_j\ket{\phi_j}\bra{\phi_j}$.
\begin{equation}
\epsilon_{1}=\frac{1}{2}(-\Omega-\Omega_R),~~\text{and},~~\epsilon_2=\frac{1}{2}(-\Omega+\Omega_R),
\end{equation}
where, $\Delta=\omega_0-\Omega$, and $\Omega_R=\sqrt{\Delta^2+4g^2}$. Now the corresponding eigenvectors are given by,
\begin{equation}
\ket{\phi_1}=\cos\theta \ket{0}-\sin\theta\ket{1},~~\ket{\phi_2}=\sin\theta\ket{0}+\cos\theta\ket{1},
\end{equation}
where, $\sigma_z\ket{0}=-\ket{0}$, $\sigma_z\ket{1}=\ket{1}$, and $\tan 2\theta=\frac{\Delta}{\Omega_R}$. As, $\ket{\phi_j(t)}=P(t)\ket{\phi_j}$, we get,
\begin{align}
&\ket{\phi_1(t)}=\cos\theta \ket{0}-e^{-i\Omega t}\sin\theta\ket{1},\\
&\ket{\phi_2(t)}=\sin\theta\ket{0}+e^{-i\Omega t}\cos\theta\ket{1}
\end{align}
With above relations we calculate $S_{kj}^{\alpha}$. 
\begin{align}
\nonumber
&S_{kj}^0=0,~~ S_{11}^1=S_{11}^{-1}=-\frac{\sin 2\theta}{2},~~S_{22}^1=S_{22}^{-1}=\frac{\sin 2\theta}{2},\\
\nonumber
& S_{12}^1=\frac{1}{2}(\cos 2\theta-1),~~ S_{12}^{-1}=\frac{1}{2}(\cos 2\theta+1),\\
&S_{21}^1=\frac{1}{2}(\cos 2\theta+1),~~S_{21}^{-1}=\frac{1}{2}(\cos 2\theta-1)
\end{align}

Fourier coefficients other than $\alpha=\pm 1$ are zero.

\twocolumngrid
\bibliography{QSL-NM}

\begin{thebibliography}{87}%
\makeatletter
\providecommand \@ifxundefined [1]{%
 \@ifx{#1\undefined}
}%
\providecommand \@ifnum [1]{%
 \ifnum #1\expandafter \@firstoftwo
 \else \expandafter \@secondoftwo
 \fi
}%
\providecommand \@ifx [1]{%
 \ifx #1\expandafter \@firstoftwo
 \else \expandafter \@secondoftwo
 \fi
}%
\providecommand \natexlab [1]{#1}%
\providecommand \enquote  [1]{``#1''}%
\providecommand \bibnamefont  [1]{#1}%
\providecommand \bibfnamefont [1]{#1}%
\providecommand \citenamefont [1]{#1}%
\providecommand \href@noop [0]{\@secondoftwo}%
\providecommand \href [0]{\begingroup \@sanitize@url \@href}%
\providecommand \@href[1]{\@@startlink{#1}\@@href}%
\providecommand \@@href[1]{\endgroup#1\@@endlink}%
\providecommand \@sanitize@url [0]{\catcode `\\12\catcode `\$12\catcode
  `\&12\catcode `\#12\catcode `\^12\catcode `\_12\catcode `\%12\relax}%
\providecommand \@@startlink[1]{}%
\providecommand \@@endlink[0]{}%
\providecommand \url  [0]{\begingroup\@sanitize@url \@url }%
\providecommand \@url [1]{\endgroup\@href {#1}{\urlprefix }}%
\providecommand \urlprefix  [0]{URL }%
\providecommand \Eprint [0]{\href }%
\providecommand \doibase [0]{https://doi.org/}%
\providecommand \selectlanguage [0]{\@gobble}%
\providecommand \bibinfo  [0]{\@secondoftwo}%
\providecommand \bibfield  [0]{\@secondoftwo}%
\providecommand \translation [1]{[#1]}%
\providecommand \BibitemOpen [0]{}%
\providecommand \bibitemStop [0]{}%
\providecommand \bibitemNoStop [0]{.\EOS\space}%
\providecommand \EOS [0]{\spacefactor3000\relax}%
\providecommand \BibitemShut  [1]{\csname bibitem#1\endcsname}%
\let\auto@bib@innerbib\@empty
%</preamble>
\bibitem [{\citenamefont {Kosloff}(2013)}]{kosloff13quantum}%
  \BibitemOpen
  \bibfield  {author} {\bibinfo {author} {\bibfnamefont {R.}~\bibnamefont
  {Kosloff}},\ }\bibfield  {title} {\bibinfo {title} {Quantum thermodynamics: A
  dynamical viewpoint},\ }\href {https://doi.org/10.3390/e15062100} {\bibfield
  {journal} {\bibinfo  {journal} {Entropy}\ }\textbf {\bibinfo {volume} {15}},\
  \bibinfo {pages} {2100} (\bibinfo {year} {2013})}\BibitemShut {NoStop}%
\bibitem [{\citenamefont {Ghosh}\ \emph {et~al.}(2018)\citenamefont {Ghosh},
  \citenamefont {Niedenzu}, \citenamefont {Mukherjee},\ and\ \citenamefont
  {Kurizki}}]{ghosh2018thermodynamic}%
  \BibitemOpen
  \bibfield  {author} {\bibinfo {author} {\bibfnamefont {A.}~\bibnamefont
  {Ghosh}}, \bibinfo {author} {\bibfnamefont {W.}~\bibnamefont {Niedenzu}},
  \bibinfo {author} {\bibfnamefont {V.}~\bibnamefont {Mukherjee}},\ and\
  \bibinfo {author} {\bibfnamefont {G.}~\bibnamefont {Kurizki}},\ }\bibinfo
  {title} {Thermodynamic principles and implementations of quantum machines},\
  in\ \href {https://doi.org/https://doi.org/10.1007/978-3-319-99046-0_2}
  {\emph {\bibinfo {booktitle} {Thermodynamics in the Quantum Regime:
  Fundamental Aspects and New Directions}}},\ \bibinfo {editor} {edited by\
  \bibinfo {editor} {\bibfnamefont {F.}~\bibnamefont {Binder}}, \bibinfo
  {editor} {\bibfnamefont {L.~A.}\ \bibnamefont {Correa}}, \bibinfo {editor}
  {\bibfnamefont {C.}~\bibnamefont {Gogolin}}, \bibinfo {editor} {\bibfnamefont
  {J.}~\bibnamefont {Anders}},\ and\ \bibinfo {editor} {\bibfnamefont
  {G.}~\bibnamefont {Adesso}}}\ (\bibinfo  {publisher} {Springer International
  Publishing},\ \bibinfo {address} {Cham},\ \bibinfo {year} {2018})\ pp.\
  \bibinfo {pages} {37--66}\BibitemShut {NoStop}%
\bibitem [{\citenamefont {Vinjanampathy}\ and\ \citenamefont
  {Anders}(2016)}]{vinjanampathy16quantum}%
  \BibitemOpen
  \bibfield  {author} {\bibinfo {author} {\bibfnamefont {S.}~\bibnamefont
  {Vinjanampathy}}\ and\ \bibinfo {author} {\bibfnamefont {J.}~\bibnamefont
  {Anders}},\ }\bibfield  {title} {\bibinfo {title} {Quantum thermodynamics},\
  }\href {https://doi.org/10.1080/00107514.2016.1201896} {\bibfield  {journal}
  {\bibinfo  {journal} {Contemporary Physics}\ }\textbf {\bibinfo {volume}
  {57}},\ \bibinfo {pages} {545} (\bibinfo {year} {2016})}\BibitemShut
  {NoStop}%
\bibitem [{\citenamefont {Esposito}\ \emph {et~al.}(2009)\citenamefont
  {Esposito}, \citenamefont {Harbola},\ and\ \citenamefont
  {Mukamel}}]{esposito-fluctuation-review}%
  \BibitemOpen
  \bibfield  {author} {\bibinfo {author} {\bibfnamefont {M.}~\bibnamefont
  {Esposito}}, \bibinfo {author} {\bibfnamefont {U.}~\bibnamefont {Harbola}},\
  and\ \bibinfo {author} {\bibfnamefont {S.}~\bibnamefont {Mukamel}},\
  }\bibfield  {title} {\bibinfo {title} {Nonequilibrium fluctuations,
  fluctuation theorems, and counting statistics in quantum systems},\ }\href
  {https://doi.org/10.1103/RevModPhys.81.1665} {\bibfield  {journal} {\bibinfo
  {journal} {Rev. Mod. Phys.}\ }\textbf {\bibinfo {volume} {81}},\ \bibinfo
  {pages} {1665} (\bibinfo {year} {2009})}\BibitemShut {NoStop}%
\bibitem [{\citenamefont {Campisi}\ \emph {et~al.}(2011)\citenamefont
  {Campisi}, \citenamefont {H\"anggi},\ and\ \citenamefont
  {Talkner}}]{campisi-fluctuation-review}%
  \BibitemOpen
  \bibfield  {author} {\bibinfo {author} {\bibfnamefont {M.}~\bibnamefont
  {Campisi}}, \bibinfo {author} {\bibfnamefont {P.}~\bibnamefont {H\"anggi}},\
  and\ \bibinfo {author} {\bibfnamefont {P.}~\bibnamefont {Talkner}},\
  }\bibfield  {title} {\bibinfo {title} {Colloquium: Quantum fluctuation
  relations: Foundations and applications},\ }\href
  {https://doi.org/10.1103/RevModPhys.83.771} {\bibfield  {journal} {\bibinfo
  {journal} {Rev. Mod. Phys.}\ }\textbf {\bibinfo {volume} {83}},\ \bibinfo
  {pages} {771} (\bibinfo {year} {2011})}\BibitemShut {NoStop}%
\bibitem [{\citenamefont {Jarzynski}(2011)}]{annurev-jarzynski}%
  \BibitemOpen
  \bibfield  {author} {\bibinfo {author} {\bibfnamefont {C.}~\bibnamefont
  {Jarzynski}},\ }\bibfield  {title} {\bibinfo {title} {Equalities and
  inequalities: Irreversibility and the second law of thermodynamics at the
  nanoscale},\ }\href
  {https://doi.org/10.1146/annurev-conmatphys-062910-140506} {\bibfield
  {journal} {\bibinfo  {journal} {Annual Review of Condensed Matter Physics}\
  }\textbf {\bibinfo {volume} {2}},\ \bibinfo {pages} {329} (\bibinfo {year}
  {2011})},\ \Eprint
  {https://arxiv.org/abs/https://doi.org/10.1146/annurev-conmatphys-062910-140506}
  {https://doi.org/10.1146/annurev-conmatphys-062910-140506} \BibitemShut
  {NoStop}%
\bibitem [{\citenamefont {Funo}\ \emph {et~al.}(2018)\citenamefont {Funo},
  \citenamefont {Ueda},\ and\ \citenamefont {Sagawa}}]{funo2018}%
  \BibitemOpen
  \bibfield  {author} {\bibinfo {author} {\bibfnamefont {K.}~\bibnamefont
  {Funo}}, \bibinfo {author} {\bibfnamefont {M.}~\bibnamefont {Ueda}},\ and\
  \bibinfo {author} {\bibfnamefont {T.}~\bibnamefont {Sagawa}},\ }\bibinfo
  {title} {Quantum fluctuation theorems},\ in\ \href
  {https://doi.org/10.1007/978-3-319-99046-0_10} {\emph {\bibinfo {booktitle}
  {Thermodynamics in the Quantum Regime: Fundamental Aspects and New
  Directions}}},\ \bibinfo {editor} {edited by\ \bibinfo {editor}
  {\bibfnamefont {F.}~\bibnamefont {Binder}}, \bibinfo {editor} {\bibfnamefont
  {L.~A.}\ \bibnamefont {Correa}}, \bibinfo {editor} {\bibfnamefont
  {C.}~\bibnamefont {Gogolin}}, \bibinfo {editor} {\bibfnamefont
  {J.}~\bibnamefont {Anders}},\ and\ \bibinfo {editor} {\bibfnamefont
  {G.}~\bibnamefont {Adesso}}}\ (\bibinfo  {publisher} {Springer International
  Publishing},\ \bibinfo {address} {Cham},\ \bibinfo {year} {2018})\ pp.\
  \bibinfo {pages} {249--273}\BibitemShut {NoStop}%
\bibitem [{\citenamefont {Barato}\ and\ \citenamefont
  {Seifert}(2015)}]{TUR1st}%
  \BibitemOpen
  \bibfield  {author} {\bibinfo {author} {\bibfnamefont {A.~C.}\ \bibnamefont
  {Barato}}\ and\ \bibinfo {author} {\bibfnamefont {U.}~\bibnamefont
  {Seifert}},\ }\bibfield  {title} {\bibinfo {title} {Thermodynamic uncertainty
  relation for biomolecular processes},\ }\href
  {https://doi.org/10.1103/PhysRevLett.114.158101} {\bibfield  {journal}
  {\bibinfo  {journal} {Phys. Rev. Lett.}\ }\textbf {\bibinfo {volume} {114}},\
  \bibinfo {pages} {158101} (\bibinfo {year} {2015})}\BibitemShut {NoStop}%
\bibitem [{\citenamefont {Pietzonka}\ \emph {et~al.}(2016)\citenamefont
  {Pietzonka}, \citenamefont {Barato},\ and\ \citenamefont
  {Seifert}}]{seifert-second}%
  \BibitemOpen
  \bibfield  {author} {\bibinfo {author} {\bibfnamefont {P.}~\bibnamefont
  {Pietzonka}}, \bibinfo {author} {\bibfnamefont {A.~C.}\ \bibnamefont
  {Barato}},\ and\ \bibinfo {author} {\bibfnamefont {U.}~\bibnamefont
  {Seifert}},\ }\bibfield  {title} {\bibinfo {title} {Universal bounds on
  current fluctuations},\ }\href {https://doi.org/10.1103/PhysRevE.93.052145}
  {\bibfield  {journal} {\bibinfo  {journal} {Phys. Rev. E}\ }\textbf {\bibinfo
  {volume} {93}},\ \bibinfo {pages} {052145} (\bibinfo {year}
  {2016})}\BibitemShut {NoStop}%
\bibitem [{\citenamefont {Gingrich}\ \emph {et~al.}(2016)\citenamefont
  {Gingrich}, \citenamefont {Horowitz}, \citenamefont {Perunov},\ and\
  \citenamefont {England}}]{TUR-Gingrich}%
  \BibitemOpen
  \bibfield  {author} {\bibinfo {author} {\bibfnamefont {T.~R.}\ \bibnamefont
  {Gingrich}}, \bibinfo {author} {\bibfnamefont {J.~M.}\ \bibnamefont
  {Horowitz}}, \bibinfo {author} {\bibfnamefont {N.}~\bibnamefont {Perunov}},\
  and\ \bibinfo {author} {\bibfnamefont {J.~L.}\ \bibnamefont {England}},\
  }\bibfield  {title} {\bibinfo {title} {Dissipation bounds all steady-state
  current fluctuations},\ }\href
  {https://doi.org/10.1103/PhysRevLett.116.120601} {\bibfield  {journal}
  {\bibinfo  {journal} {Phys. Rev. Lett.}\ }\textbf {\bibinfo {volume} {116}},\
  \bibinfo {pages} {120601} (\bibinfo {year} {2016})}\BibitemShut {NoStop}%
\bibitem [{\citenamefont {Pietzonka}\ \emph {et~al.}(2017)\citenamefont
  {Pietzonka}, \citenamefont {Ritort},\ and\ \citenamefont
  {Seifert}}]{Pietzonka-1}%
  \BibitemOpen
  \bibfield  {author} {\bibinfo {author} {\bibfnamefont {P.}~\bibnamefont
  {Pietzonka}}, \bibinfo {author} {\bibfnamefont {F.}~\bibnamefont {Ritort}},\
  and\ \bibinfo {author} {\bibfnamefont {U.}~\bibnamefont {Seifert}},\
  }\bibfield  {title} {\bibinfo {title} {Finite-time generalization of the
  thermodynamic uncertainty relation},\ }\href
  {https://doi.org/10.1103/PhysRevE.96.012101} {\bibfield  {journal} {\bibinfo
  {journal} {Phys. Rev. E}\ }\textbf {\bibinfo {volume} {96}},\ \bibinfo
  {pages} {012101} (\bibinfo {year} {2017})}\BibitemShut {NoStop}%
\bibitem [{\citenamefont {Horowitz}\ and\ \citenamefont
  {Gingrich}(2017)}]{horowitz-finite}%
  \BibitemOpen
  \bibfield  {author} {\bibinfo {author} {\bibfnamefont {J.~M.}\ \bibnamefont
  {Horowitz}}\ and\ \bibinfo {author} {\bibfnamefont {T.~R.}\ \bibnamefont
  {Gingrich}},\ }\bibfield  {title} {\bibinfo {title} {Proof of the finite-time
  thermodynamic uncertainty relation for steady-state currents},\ }\href
  {https://doi.org/10.1103/PhysRevE.96.020103} {\bibfield  {journal} {\bibinfo
  {journal} {Phys. Rev. E}\ }\textbf {\bibinfo {volume} {96}},\ \bibinfo
  {pages} {020103} (\bibinfo {year} {2017})}\BibitemShut {NoStop}%
\bibitem [{\citenamefont {Pietzonka}\ and\ \citenamefont
  {Seifert}(2018)}]{TUR-tradeoff}%
  \BibitemOpen
  \bibfield  {author} {\bibinfo {author} {\bibfnamefont {P.}~\bibnamefont
  {Pietzonka}}\ and\ \bibinfo {author} {\bibfnamefont {U.}~\bibnamefont
  {Seifert}},\ }\bibfield  {title} {\bibinfo {title} {Universal trade-off
  between power, efficiency, and constancy in steady-state heat engines},\
  }\href {https://doi.org/10.1103/PhysRevLett.120.190602} {\bibfield  {journal}
  {\bibinfo  {journal} {Phys. Rev. Lett.}\ }\textbf {\bibinfo {volume} {120}},\
  \bibinfo {pages} {190602} (\bibinfo {year} {2018})}\BibitemShut {NoStop}%
\bibitem [{\citenamefont {Horowitz}\ and\ \citenamefont
  {Gingrich}(2020)}]{tur-satis3}%
  \BibitemOpen
  \bibfield  {author} {\bibinfo {author} {\bibfnamefont {J.~M.}\ \bibnamefont
  {Horowitz}}\ and\ \bibinfo {author} {\bibfnamefont {T.~R.}\ \bibnamefont
  {Gingrich}},\ }\bibfield  {title} {\bibinfo {title} {Thermodynamic
  uncertainty relations constrain non-equilibrium fluctuations},\ }\href
  {https://doi.org/https://doi.org/10.1038/s41567-019-0702-6} {\bibfield
  {journal} {\bibinfo  {journal} {Nat. Phys.}\ }\textbf {\bibinfo {volume}
  {16}},\ \bibinfo {pages} {15} (\bibinfo {year} {2020})}\BibitemShut {NoStop}%
\bibitem [{\citenamefont {Polettini}\ \emph {et~al.}(2016)\citenamefont
  {Polettini}, \citenamefont {Lazarescu},\ and\ \citenamefont
  {Esposito}}]{polettini-1}%
  \BibitemOpen
  \bibfield  {author} {\bibinfo {author} {\bibfnamefont {M.}~\bibnamefont
  {Polettini}}, \bibinfo {author} {\bibfnamefont {A.}~\bibnamefont
  {Lazarescu}},\ and\ \bibinfo {author} {\bibfnamefont {M.}~\bibnamefont
  {Esposito}},\ }\bibfield  {title} {\bibinfo {title} {Tightening the
  uncertainty principle for stochastic currents},\ }\href
  {https://doi.org/10.1103/PhysRevE.94.052104} {\bibfield  {journal} {\bibinfo
  {journal} {Phys. Rev. E}\ }\textbf {\bibinfo {volume} {94}},\ \bibinfo
  {pages} {052104} (\bibinfo {year} {2016})}\BibitemShut {NoStop}%
\bibitem [{\citenamefont {Gingrich}\ \emph {et~al.}(2017)\citenamefont
  {Gingrich}, \citenamefont {Rotskoff},\ and\ \citenamefont
  {Horowitz}}]{Gingrich_2017}%
  \BibitemOpen
  \bibfield  {author} {\bibinfo {author} {\bibfnamefont {T.~R.}\ \bibnamefont
  {Gingrich}}, \bibinfo {author} {\bibfnamefont {G.~M.}\ \bibnamefont
  {Rotskoff}},\ and\ \bibinfo {author} {\bibfnamefont {J.~M.}\ \bibnamefont
  {Horowitz}},\ }\bibfield  {title} {\bibinfo {title} {Inferring dissipation
  from current fluctuations},\ }\href
  {https://doi.org/10.1088/1751-8121/aa672f} {\bibfield  {journal} {\bibinfo
  {journal} {Journal of Physics A: Mathematical and Theoretical}\ }\textbf
  {\bibinfo {volume} {50}},\ \bibinfo {pages} {184004} (\bibinfo {year}
  {2017})}\BibitemShut {NoStop}%
\bibitem [{\citenamefont {Dechant}\ and\ \citenamefont {ichi
  Sasa}(2018)}]{Dechant_2018}%
  \BibitemOpen
  \bibfield  {author} {\bibinfo {author} {\bibfnamefont {A.}~\bibnamefont
  {Dechant}}\ and\ \bibinfo {author} {\bibfnamefont {S.}~\bibnamefont {ichi
  Sasa}},\ }\bibfield  {title} {\bibinfo {title} {Current fluctuations and
  transport efficiency for general langevin systems},\ }\href
  {https://doi.org/10.1088/1742-5468/aac91a} {\bibfield  {journal} {\bibinfo
  {journal} {Journal of Statistical Mechanics: Theory and Experiment}\ }\textbf
  {\bibinfo {volume} {2018}},\ \bibinfo {pages} {063209} (\bibinfo {year}
  {2018})}\BibitemShut {NoStop}%
\bibitem [{\citenamefont {Proesmans}\ and\ \citenamefont
  {Horowitz}(2019)}]{Proesmans_2019}%
  \BibitemOpen
  \bibfield  {author} {\bibinfo {author} {\bibfnamefont {K.}~\bibnamefont
  {Proesmans}}\ and\ \bibinfo {author} {\bibfnamefont {J.~M.}\ \bibnamefont
  {Horowitz}},\ }\bibfield  {title} {\bibinfo {title} {Hysteretic thermodynamic
  uncertainty relation for systems with broken time-reversal symmetry},\ }\href
  {https://doi.org/10.1088/1742-5468/ab14da} {\bibfield  {journal} {\bibinfo
  {journal} {Journal of Statistical Mechanics: Theory and Experiment}\ }\textbf
  {\bibinfo {volume} {2019}},\ \bibinfo {pages} {054005} (\bibinfo {year}
  {2019})}\BibitemShut {NoStop}%
\bibitem [{\citenamefont {{Shiraishi}}(2017)}]{naoto-discrete}%
  \BibitemOpen
  \bibfield  {author} {\bibinfo {author} {\bibfnamefont {N.}~\bibnamefont
  {{Shiraishi}}},\ }\bibfield  {title} {\bibinfo {title} {{Finite-time
  thermodynamic uncertainty relation do not hold for discrete-time Markov
  process}},\ }\href@noop {} {\bibfield  {journal} {\bibinfo  {journal} {arXiv
  e-prints}\ ,\ \bibinfo {eid} {arXiv:1706.00892}} (\bibinfo {year} {2017})},\
  \Eprint {https://arxiv.org/abs/1706.00892} {arXiv:1706.00892
  [cond-mat.stat-mech]} \BibitemShut {NoStop}%
\bibitem [{\citenamefont {Proesmans}\ and\ \citenamefont {den
  Broeck}(2017)}]{Proesmans_2017}%
  \BibitemOpen
  \bibfield  {author} {\bibinfo {author} {\bibfnamefont {K.}~\bibnamefont
  {Proesmans}}\ and\ \bibinfo {author} {\bibfnamefont {C.~V.}\ \bibnamefont
  {den Broeck}},\ }\bibfield  {title} {\bibinfo {title} {Discrete-time
  thermodynamic uncertainty relation},\ }\href
  {https://doi.org/10.1209/0295-5075/119/20001} {\bibfield  {journal} {\bibinfo
   {journal} {{EPL} (Europhysics Letters)}\ }\textbf {\bibinfo {volume}
  {119}},\ \bibinfo {pages} {20001} (\bibinfo {year} {2017})}\BibitemShut
  {NoStop}%
\bibitem [{\citenamefont {Chiuchi\`u}\ and\ \citenamefont
  {Pigolotti}(2018)}]{discrete-3}%
  \BibitemOpen
  \bibfield  {author} {\bibinfo {author} {\bibfnamefont {D.}~\bibnamefont
  {Chiuchi\`u}}\ and\ \bibinfo {author} {\bibfnamefont {S.}~\bibnamefont
  {Pigolotti}},\ }\bibfield  {title} {\bibinfo {title} {Mapping of uncertainty
  relations between continuous and discrete time},\ }\href
  {https://doi.org/10.1103/PhysRevE.97.032109} {\bibfield  {journal} {\bibinfo
  {journal} {Phys. Rev. E}\ }\textbf {\bibinfo {volume} {97}},\ \bibinfo
  {pages} {032109} (\bibinfo {year} {2018})}\BibitemShut {NoStop}%
\bibitem [{\citenamefont {Barato}\ and\ \citenamefont
  {Seifert}(2016)}]{barato-clock}%
  \BibitemOpen
  \bibfield  {author} {\bibinfo {author} {\bibfnamefont {A.~C.}\ \bibnamefont
  {Barato}}\ and\ \bibinfo {author} {\bibfnamefont {U.}~\bibnamefont
  {Seifert}},\ }\bibfield  {title} {\bibinfo {title} {Cost and precision of
  brownian clocks},\ }\href {https://doi.org/10.1103/PhysRevX.6.041053}
  {\bibfield  {journal} {\bibinfo  {journal} {Phys. Rev. X}\ }\textbf {\bibinfo
  {volume} {6}},\ \bibinfo {pages} {041053} (\bibinfo {year}
  {2016})}\BibitemShut {NoStop}%
\bibitem [{\citenamefont {Barato}\ \emph {et~al.}(2018)\citenamefont {Barato},
  \citenamefont {Chetrite}, \citenamefont {Faggionato},\ and\ \citenamefont
  {Gabrielli}}]{Barato_2018}%
  \BibitemOpen
  \bibfield  {author} {\bibinfo {author} {\bibfnamefont {A.~C.}\ \bibnamefont
  {Barato}}, \bibinfo {author} {\bibfnamefont {R.}~\bibnamefont {Chetrite}},
  \bibinfo {author} {\bibfnamefont {A.}~\bibnamefont {Faggionato}},\ and\
  \bibinfo {author} {\bibfnamefont {D.}~\bibnamefont {Gabrielli}},\ }\bibfield
  {title} {\bibinfo {title} {Bounds on current fluctuations in periodically
  driven systems},\ }\href {https://doi.org/10.1088/1367-2630/aae512}
  {\bibfield  {journal} {\bibinfo  {journal} {New Journal of Physics}\ }\textbf
  {\bibinfo {volume} {20}},\ \bibinfo {pages} {103023} (\bibinfo {year}
  {2018})}\BibitemShut {NoStop}%
\bibitem [{\citenamefont {Koyuk}\ \emph {et~al.}(2018)\citenamefont {Koyuk},
  \citenamefont {Seifert},\ and\ \citenamefont {Pietzonka}}]{Koyuk_2018}%
  \BibitemOpen
  \bibfield  {author} {\bibinfo {author} {\bibfnamefont {T.}~\bibnamefont
  {Koyuk}}, \bibinfo {author} {\bibfnamefont {U.}~\bibnamefont {Seifert}},\
  and\ \bibinfo {author} {\bibfnamefont {P.}~\bibnamefont {Pietzonka}},\
  }\bibfield  {title} {\bibinfo {title} {A generalization of the thermodynamic
  uncertainty relation to periodically driven systems},\ }\href
  {https://doi.org/10.1088/1751-8121/aaeec4} {\bibfield  {journal} {\bibinfo
  {journal} {Journal of Physics A: Mathematical and Theoretical}\ }\textbf
  {\bibinfo {volume} {52}},\ \bibinfo {pages} {02LT02} (\bibinfo {year}
  {2018})}\BibitemShut {NoStop}%
\bibitem [{\citenamefont {Dechant}\ and\ \citenamefont {Sasa}(2018)}]{sasa-2}%
  \BibitemOpen
  \bibfield  {author} {\bibinfo {author} {\bibfnamefont {A.}~\bibnamefont
  {Dechant}}\ and\ \bibinfo {author} {\bibfnamefont {S.-i.}\ \bibnamefont
  {Sasa}},\ }\bibfield  {title} {\bibinfo {title} {Entropic bounds on currents
  in langevin systems},\ }\href {https://doi.org/10.1103/PhysRevE.97.062101}
  {\bibfield  {journal} {\bibinfo  {journal} {Phys. Rev. E}\ }\textbf {\bibinfo
  {volume} {97}},\ \bibinfo {pages} {062101} (\bibinfo {year}
  {2018})}\BibitemShut {NoStop}%
\bibitem [{\citenamefont {Holubec}\ and\ \citenamefont
  {Ryabov}(2018)}]{holubec-cycling}%
  \BibitemOpen
  \bibfield  {author} {\bibinfo {author} {\bibfnamefont {V.}~\bibnamefont
  {Holubec}}\ and\ \bibinfo {author} {\bibfnamefont {A.}~\bibnamefont
  {Ryabov}},\ }\bibfield  {title} {\bibinfo {title} {Cycling tames power
  fluctuations near optimum efficiency},\ }\href
  {https://doi.org/10.1103/PhysRevLett.121.120601} {\bibfield  {journal}
  {\bibinfo  {journal} {Phys. Rev. Lett.}\ }\textbf {\bibinfo {volume} {121}},\
  \bibinfo {pages} {120601} (\bibinfo {year} {2018})}\BibitemShut {NoStop}%
\bibitem [{\citenamefont {Macieszczak}\ \emph {et~al.}(2018)\citenamefont
  {Macieszczak}, \citenamefont {Brandner},\ and\ \citenamefont
  {Garrahan}}]{TUR-Brandner}%
  \BibitemOpen
  \bibfield  {author} {\bibinfo {author} {\bibfnamefont {K.}~\bibnamefont
  {Macieszczak}}, \bibinfo {author} {\bibfnamefont {K.}~\bibnamefont
  {Brandner}},\ and\ \bibinfo {author} {\bibfnamefont {J.~P.}\ \bibnamefont
  {Garrahan}},\ }\bibfield  {title} {\bibinfo {title} {Unified thermodynamic
  uncertainty relations in linear response},\ }\href
  {https://doi.org/10.1103/PhysRevLett.121.130601} {\bibfield  {journal}
  {\bibinfo  {journal} {Phys. Rev. Lett.}\ }\textbf {\bibinfo {volume} {121}},\
  \bibinfo {pages} {130601} (\bibinfo {year} {2018})}\BibitemShut {NoStop}%
\bibitem [{\citenamefont {Barato}\ \emph {et~al.}(2019)\citenamefont {Barato},
  \citenamefont {Chetrite}, \citenamefont {Faggionato},\ and\ \citenamefont
  {Gabrielli}}]{Barato_2019}%
  \BibitemOpen
  \bibfield  {author} {\bibinfo {author} {\bibfnamefont {A.~C.}\ \bibnamefont
  {Barato}}, \bibinfo {author} {\bibfnamefont {R.}~\bibnamefont {Chetrite}},
  \bibinfo {author} {\bibfnamefont {A.}~\bibnamefont {Faggionato}},\ and\
  \bibinfo {author} {\bibfnamefont {D.}~\bibnamefont {Gabrielli}},\ }\bibfield
  {title} {\bibinfo {title} {A unifying picture of generalized thermodynamic
  uncertainty relations},\ }\href {https://doi.org/10.1088/1742-5468/ab3457}
  {\bibfield  {journal} {\bibinfo  {journal} {Journal of Statistical Mechanics:
  Theory and Experiment}\ }\textbf {\bibinfo {volume} {2019}},\ \bibinfo
  {pages} {084017} (\bibinfo {year} {2019})}\BibitemShut {NoStop}%
\bibitem [{\citenamefont {Koyuk}\ and\ \citenamefont
  {Seifert}(2020)}]{koyuk-2020}%
  \BibitemOpen
  \bibfield  {author} {\bibinfo {author} {\bibfnamefont {T.}~\bibnamefont
  {Koyuk}}\ and\ \bibinfo {author} {\bibfnamefont {U.}~\bibnamefont
  {Seifert}},\ }\bibfield  {title} {\bibinfo {title} {Thermodynamic uncertainty
  relation for time-dependent driving},\ }\href
  {https://doi.org/10.1103/PhysRevLett.125.260604} {\bibfield  {journal}
  {\bibinfo  {journal} {Phys. Rev. Lett.}\ }\textbf {\bibinfo {volume} {125}},\
  \bibinfo {pages} {260604} (\bibinfo {year} {2020})}\BibitemShut {NoStop}%
\bibitem [{\citenamefont {Cangemi}\ \emph {et~al.}(2021)\citenamefont
  {Cangemi}, \citenamefont {Carrega}, \citenamefont {De~Candia}, \citenamefont
  {Cataudella}, \citenamefont {De~Filippis}, \citenamefont {Sassetti},\ and\
  \citenamefont {Benenti}}]{benenti-2021}%
  \BibitemOpen
  \bibfield  {author} {\bibinfo {author} {\bibfnamefont {L.~M.}\ \bibnamefont
  {Cangemi}}, \bibinfo {author} {\bibfnamefont {M.}~\bibnamefont {Carrega}},
  \bibinfo {author} {\bibfnamefont {A.}~\bibnamefont {De~Candia}}, \bibinfo
  {author} {\bibfnamefont {V.}~\bibnamefont {Cataudella}}, \bibinfo {author}
  {\bibfnamefont {G.}~\bibnamefont {De~Filippis}}, \bibinfo {author}
  {\bibfnamefont {M.}~\bibnamefont {Sassetti}},\ and\ \bibinfo {author}
  {\bibfnamefont {G.}~\bibnamefont {Benenti}},\ }\bibfield  {title} {\bibinfo
  {title} {Optimal energy conversion through antiadiabatic driving breaking
  time-reversal symmetry},\ }\href
  {https://doi.org/10.1103/PhysRevResearch.3.013237} {\bibfield  {journal}
  {\bibinfo  {journal} {Phys. Rev. Research}\ }\textbf {\bibinfo {volume}
  {3}},\ \bibinfo {pages} {013237} (\bibinfo {year} {2021})}\BibitemShut
  {NoStop}%
\bibitem [{\citenamefont {Van~Vu}\ and\ \citenamefont
  {Hasegawa}(2020)}]{vu-under1}%
  \BibitemOpen
  \bibfield  {author} {\bibinfo {author} {\bibfnamefont {T.}~\bibnamefont
  {Van~Vu}}\ and\ \bibinfo {author} {\bibfnamefont {Y.}~\bibnamefont
  {Hasegawa}},\ }\bibfield  {title} {\bibinfo {title} {Thermodynamic
  uncertainty relations under arbitrary control protocols},\ }\href
  {https://doi.org/10.1103/PhysRevResearch.2.013060} {\bibfield  {journal}
  {\bibinfo  {journal} {Phys. Rev. Research}\ }\textbf {\bibinfo {volume}
  {2}},\ \bibinfo {pages} {013060} (\bibinfo {year} {2020})}\BibitemShut
  {NoStop}%
\bibitem [{\citenamefont {Van~Vu}\ and\ \citenamefont
  {Hasegawa}(2019)}]{vu-under2}%
  \BibitemOpen
  \bibfield  {author} {\bibinfo {author} {\bibfnamefont {T.}~\bibnamefont
  {Van~Vu}}\ and\ \bibinfo {author} {\bibfnamefont {Y.}~\bibnamefont
  {Hasegawa}},\ }\bibfield  {title} {\bibinfo {title} {Uncertainty relations
  for underdamped langevin dynamics},\ }\href
  {https://doi.org/10.1103/PhysRevE.100.032130} {\bibfield  {journal} {\bibinfo
   {journal} {Phys. Rev. E}\ }\textbf {\bibinfo {volume} {100}},\ \bibinfo
  {pages} {032130} (\bibinfo {year} {2019})}\BibitemShut {NoStop}%
\bibitem [{\citenamefont {Chun}\ \emph {et~al.}(2019)\citenamefont {Chun},
  \citenamefont {Fischer},\ and\ \citenamefont {Seifert}}]{udo-under}%
  \BibitemOpen
  \bibfield  {author} {\bibinfo {author} {\bibfnamefont {H.-M.}\ \bibnamefont
  {Chun}}, \bibinfo {author} {\bibfnamefont {L.~P.}\ \bibnamefont {Fischer}},\
  and\ \bibinfo {author} {\bibfnamefont {U.}~\bibnamefont {Seifert}},\
  }\bibfield  {title} {\bibinfo {title} {Effect of a magnetic field on the
  thermodynamic uncertainty relation},\ }\href
  {https://doi.org/10.1103/PhysRevE.99.042128} {\bibfield  {journal} {\bibinfo
  {journal} {Phys. Rev. E}\ }\textbf {\bibinfo {volume} {99}},\ \bibinfo
  {pages} {042128} (\bibinfo {year} {2019})}\BibitemShut {NoStop}%
\bibitem [{\citenamefont {Lee}\ \emph {et~al.}(2019)\citenamefont {Lee},
  \citenamefont {Park},\ and\ \citenamefont {Park}}]{park-under}%
  \BibitemOpen
  \bibfield  {author} {\bibinfo {author} {\bibfnamefont {J.~S.}\ \bibnamefont
  {Lee}}, \bibinfo {author} {\bibfnamefont {J.-M.}\ \bibnamefont {Park}},\ and\
  \bibinfo {author} {\bibfnamefont {H.}~\bibnamefont {Park}},\ }\bibfield
  {title} {\bibinfo {title} {Thermodynamic uncertainty relation for underdamped
  langevin systems driven by a velocity-dependent force},\ }\href
  {https://doi.org/10.1103/PhysRevE.100.062132} {\bibfield  {journal} {\bibinfo
   {journal} {Phys. Rev. E}\ }\textbf {\bibinfo {volume} {100}},\ \bibinfo
  {pages} {062132} (\bibinfo {year} {2019})}\BibitemShut {NoStop}%
\bibitem [{\citenamefont {Brandner}\ \emph {et~al.}(2018)\citenamefont
  {Brandner}, \citenamefont {Hanazato},\ and\ \citenamefont
  {Saito}}]{brander-transport}%
  \BibitemOpen
  \bibfield  {author} {\bibinfo {author} {\bibfnamefont {K.}~\bibnamefont
  {Brandner}}, \bibinfo {author} {\bibfnamefont {T.}~\bibnamefont {Hanazato}},\
  and\ \bibinfo {author} {\bibfnamefont {K.}~\bibnamefont {Saito}},\ }\bibfield
   {title} {\bibinfo {title} {Thermodynamic bounds on precision in ballistic
  multiterminal transport},\ }\href
  {https://doi.org/10.1103/PhysRevLett.120.090601} {\bibfield  {journal}
  {\bibinfo  {journal} {Phys. Rev. Lett.}\ }\textbf {\bibinfo {volume} {120}},\
  \bibinfo {pages} {090601} (\bibinfo {year} {2018})}\BibitemShut {NoStop}%
\bibitem [{\citenamefont {Potts}\ and\ \citenamefont
  {Samuelsson}(2019)}]{potts-feedback}%
  \BibitemOpen
  \bibfield  {author} {\bibinfo {author} {\bibfnamefont {P.~P.}\ \bibnamefont
  {Potts}}\ and\ \bibinfo {author} {\bibfnamefont {P.}~\bibnamefont
  {Samuelsson}},\ }\bibfield  {title} {\bibinfo {title} {Thermodynamic
  uncertainty relations including measurement and feedback},\ }\href
  {https://doi.org/10.1103/PhysRevE.100.052137} {\bibfield  {journal} {\bibinfo
   {journal} {Phys. Rev. E}\ }\textbf {\bibinfo {volume} {100}},\ \bibinfo
  {pages} {052137} (\bibinfo {year} {2019})}\BibitemShut {NoStop}%
\bibitem [{\citenamefont {Liu}\ \emph {et~al.}(2020)\citenamefont {Liu},
  \citenamefont {Gong},\ and\ \citenamefont {Ueda}}]{ueda-tur-1}%
  \BibitemOpen
  \bibfield  {author} {\bibinfo {author} {\bibfnamefont {K.}~\bibnamefont
  {Liu}}, \bibinfo {author} {\bibfnamefont {Z.}~\bibnamefont {Gong}},\ and\
  \bibinfo {author} {\bibfnamefont {M.}~\bibnamefont {Ueda}},\ }\bibfield
  {title} {\bibinfo {title} {Thermodynamic uncertainty relation for arbitrary
  initial states},\ }\href {https://doi.org/10.1103/PhysRevLett.125.140602}
  {\bibfield  {journal} {\bibinfo  {journal} {Phys. Rev. Lett.}\ }\textbf
  {\bibinfo {volume} {125}},\ \bibinfo {pages} {140602} (\bibinfo {year}
  {2020})}\BibitemShut {NoStop}%
\bibitem [{\citenamefont {Agarwalla}\ and\ \citenamefont
  {Segal}(2018)}]{TURBijay2}%
  \BibitemOpen
  \bibfield  {author} {\bibinfo {author} {\bibfnamefont {B.~K.}\ \bibnamefont
  {Agarwalla}}\ and\ \bibinfo {author} {\bibfnamefont {D.}~\bibnamefont
  {Segal}},\ }\bibfield  {title} {\bibinfo {title} {Assessing the validity of
  the thermodynamic uncertainty relation in quantum systems},\ }\href
  {https://doi.org/10.1103/PhysRevB.98.155438} {\bibfield  {journal} {\bibinfo
  {journal} {Phys. Rev. B}\ }\textbf {\bibinfo {volume} {98}},\ \bibinfo
  {pages} {155438} (\bibinfo {year} {2018})}\BibitemShut {NoStop}%
\bibitem [{\citenamefont {Liu}\ and\ \citenamefont {Segal}(2019)}]{TURDvira}%
  \BibitemOpen
  \bibfield  {author} {\bibinfo {author} {\bibfnamefont {J.}~\bibnamefont
  {Liu}}\ and\ \bibinfo {author} {\bibfnamefont {D.}~\bibnamefont {Segal}},\
  }\bibfield  {title} {\bibinfo {title} {Thermodynamic uncertainty relation in
  quantum thermoelectric junctions},\ }\href
  {https://doi.org/10.1103/PhysRevE.99.062141} {\bibfield  {journal} {\bibinfo
  {journal} {Phys. Rev. E}\ }\textbf {\bibinfo {volume} {99}},\ \bibinfo
  {pages} {062141} (\bibinfo {year} {2019})}\BibitemShut {NoStop}%
\bibitem [{\citenamefont {Pal}\ \emph {et~al.}(2020)\citenamefont {Pal},
  \citenamefont {Saryal}, \citenamefont {Segal}, \citenamefont {Mahesh},\ and\
  \citenamefont {Agarwalla}}]{bijay-exp}%
  \BibitemOpen
  \bibfield  {author} {\bibinfo {author} {\bibfnamefont {S.}~\bibnamefont
  {Pal}}, \bibinfo {author} {\bibfnamefont {S.}~\bibnamefont {Saryal}},
  \bibinfo {author} {\bibfnamefont {D.}~\bibnamefont {Segal}}, \bibinfo
  {author} {\bibfnamefont {T.~S.}\ \bibnamefont {Mahesh}},\ and\ \bibinfo
  {author} {\bibfnamefont {B.~K.}\ \bibnamefont {Agarwalla}},\ }\bibfield
  {title} {\bibinfo {title} {Experimental study of the thermodynamic
  uncertainty relation},\ }\href
  {https://doi.org/10.1103/PhysRevResearch.2.022044} {\bibfield  {journal}
  {\bibinfo  {journal} {Phys. Rev. Research}\ }\textbf {\bibinfo {volume}
  {2}},\ \bibinfo {pages} {022044} (\bibinfo {year} {2020})}\BibitemShut
  {NoStop}%
\bibitem [{\citenamefont {Saryal}\ \emph
  {et~al.}(2021{\natexlab{a}})\citenamefont {Saryal}, \citenamefont {Sadekar},\
  and\ \citenamefont {Agarwalla}}]{TURBijay1}%
  \BibitemOpen
  \bibfield  {author} {\bibinfo {author} {\bibfnamefont {S.}~\bibnamefont
  {Saryal}}, \bibinfo {author} {\bibfnamefont {O.}~\bibnamefont {Sadekar}},\
  and\ \bibinfo {author} {\bibfnamefont {B.~K.}\ \bibnamefont {Agarwalla}},\
  }\bibfield  {title} {\bibinfo {title} {Thermodynamic uncertainty relation for
  energy transport in a transient regime: A model study},\ }\href
  {https://doi.org/10.1103/PhysRevE.103.022141} {\bibfield  {journal} {\bibinfo
   {journal} {Phys. Rev. E}\ }\textbf {\bibinfo {volume} {103}},\ \bibinfo
  {pages} {022141} (\bibinfo {year} {2021}{\natexlab{a}})}\BibitemShut
  {NoStop}%
\bibitem [{\citenamefont {Carollo}\ \emph {et~al.}(2019)\citenamefont
  {Carollo}, \citenamefont {Jack},\ and\ \citenamefont
  {Garrahan}}]{large-deviation-quantum}%
  \BibitemOpen
  \bibfield  {author} {\bibinfo {author} {\bibfnamefont {F.}~\bibnamefont
  {Carollo}}, \bibinfo {author} {\bibfnamefont {R.~L.}\ \bibnamefont {Jack}},\
  and\ \bibinfo {author} {\bibfnamefont {J.~P.}\ \bibnamefont {Garrahan}},\
  }\bibfield  {title} {\bibinfo {title} {Unraveling the large deviation
  statistics of markovian open quantum systems},\ }\href
  {https://doi.org/10.1103/PhysRevLett.122.130605} {\bibfield  {journal}
  {\bibinfo  {journal} {Phys. Rev. Lett.}\ }\textbf {\bibinfo {volume} {122}},\
  \bibinfo {pages} {130605} (\bibinfo {year} {2019})}\BibitemShut {NoStop}%
\bibitem [{\citenamefont {Guarnieri}\ \emph {et~al.}(2019)\citenamefont
  {Guarnieri}, \citenamefont {Landi}, \citenamefont {Clark},\ and\
  \citenamefont {Goold}}]{goold-tur}%
  \BibitemOpen
  \bibfield  {author} {\bibinfo {author} {\bibfnamefont {G.}~\bibnamefont
  {Guarnieri}}, \bibinfo {author} {\bibfnamefont {G.~T.}\ \bibnamefont
  {Landi}}, \bibinfo {author} {\bibfnamefont {S.~R.}\ \bibnamefont {Clark}},\
  and\ \bibinfo {author} {\bibfnamefont {J.}~\bibnamefont {Goold}},\ }\bibfield
   {title} {\bibinfo {title} {Thermodynamics of precision in quantum
  nonequilibrium steady states},\ }\href
  {https://doi.org/10.1103/PhysRevResearch.1.033021} {\bibfield  {journal}
  {\bibinfo  {journal} {Phys. Rev. Research}\ }\textbf {\bibinfo {volume}
  {1}},\ \bibinfo {pages} {033021} (\bibinfo {year} {2019})}\BibitemShut
  {NoStop}%
\bibitem [{\citenamefont {Hasegawa}(2020)}]{hasegawa-tur1}%
  \BibitemOpen
  \bibfield  {author} {\bibinfo {author} {\bibfnamefont {Y.}~\bibnamefont
  {Hasegawa}},\ }\bibfield  {title} {\bibinfo {title} {Quantum thermodynamic
  uncertainty relation for continuous measurement},\ }\href
  {https://doi.org/10.1103/PhysRevLett.125.050601} {\bibfield  {journal}
  {\bibinfo  {journal} {Phys. Rev. Lett.}\ }\textbf {\bibinfo {volume} {125}},\
  \bibinfo {pages} {050601} (\bibinfo {year} {2020})}\BibitemShut {NoStop}%
\bibitem [{\citenamefont {Hasegawa}(2021)}]{hasegawa-tur2}%
  \BibitemOpen
  \bibfield  {author} {\bibinfo {author} {\bibfnamefont {Y.}~\bibnamefont
  {Hasegawa}},\ }\bibfield  {title} {\bibinfo {title} {Thermodynamic
  uncertainty relation for general open quantum systems},\ }\href
  {https://doi.org/10.1103/PhysRevLett.126.010602} {\bibfield  {journal}
  {\bibinfo  {journal} {Phys. Rev. Lett.}\ }\textbf {\bibinfo {volume} {126}},\
  \bibinfo {pages} {010602} (\bibinfo {year} {2021})}\BibitemShut {NoStop}%
\bibitem [{\citenamefont {Ptaszy{\'n}ski}(2018)}]{coherent-4}%
  \BibitemOpen
  \bibfield  {author} {\bibinfo {author} {\bibfnamefont {K.}~\bibnamefont
  {Ptaszy{\'n}ski}},\ }\bibfield  {title} {\bibinfo {title} {Coherence-enhanced
  constancy of a quantum thermoelectric generator},\ }\href
  {https://doi.org/https://doi.org/10.1103/PhysRevB.98.085425} {\bibfield
  {journal} {\bibinfo  {journal} {Phys. Rev. B}\ }\textbf {\bibinfo {volume}
  {98}},\ \bibinfo {pages} {085425} (\bibinfo {year} {2018})}\BibitemShut
  {NoStop}%
\bibitem [{\citenamefont {Cangemi}\ \emph {et~al.}(2020)\citenamefont
  {Cangemi}, \citenamefont {Cataudella}, \citenamefont {Benenti}, \citenamefont
  {Sassetti},\ and\ \citenamefont {De~Filippis}}]{coherent-6}%
  \BibitemOpen
  \bibfield  {author} {\bibinfo {author} {\bibfnamefont {L.~M.}\ \bibnamefont
  {Cangemi}}, \bibinfo {author} {\bibfnamefont {V.}~\bibnamefont {Cataudella}},
  \bibinfo {author} {\bibfnamefont {G.}~\bibnamefont {Benenti}}, \bibinfo
  {author} {\bibfnamefont {M.}~\bibnamefont {Sassetti}},\ and\ \bibinfo
  {author} {\bibfnamefont {G.}~\bibnamefont {De~Filippis}},\ }\bibfield
  {title} {\bibinfo {title} {Violation of thermodynamics uncertainty relations
  in a periodically driven work-to-work converter from weak to strong
  dissipation},\ }\href {https://doi.org/10.1103/PhysRevB.102.165418}
  {\bibfield  {journal} {\bibinfo  {journal} {Phys. Rev. B}\ }\textbf {\bibinfo
  {volume} {102}},\ \bibinfo {pages} {165418} (\bibinfo {year}
  {2020})}\BibitemShut {NoStop}%
\bibitem [{\citenamefont {Kalaee}\ \emph {et~al.}(2021)\citenamefont {Kalaee},
  \citenamefont {Wacker},\ and\ \citenamefont {Potts}}]{coherent-2}%
  \BibitemOpen
  \bibfield  {author} {\bibinfo {author} {\bibfnamefont {A.~A.~S.}\
  \bibnamefont {Kalaee}}, \bibinfo {author} {\bibfnamefont {A.}~\bibnamefont
  {Wacker}},\ and\ \bibinfo {author} {\bibfnamefont {P.~P.}\ \bibnamefont
  {Potts}},\ }\bibfield  {title} {\bibinfo {title} {Violating the thermodynamic
  uncertainty relation in the three-level maser},\ }\href
  {https://doi.org/10.1103/PhysRevE.104.L012103} {\bibfield  {journal}
  {\bibinfo  {journal} {Phys. Rev. E}\ }\textbf {\bibinfo {volume} {104}},\
  \bibinfo {pages} {L012103} (\bibinfo {year} {2021})}\BibitemShut {NoStop}%
\bibitem [{\citenamefont {Singh}\ and\ \citenamefont
  {Hyeon}(2021)}]{coherent-3}%
  \BibitemOpen
  \bibfield  {author} {\bibinfo {author} {\bibfnamefont {D.}~\bibnamefont
  {Singh}}\ and\ \bibinfo {author} {\bibfnamefont {C.}~\bibnamefont {Hyeon}},\
  }\bibfield  {title} {\bibinfo {title} {Origin of loose bound of the
  thermodynamic uncertainty relation in a dissipative two-level quantum
  system},\ }\href {https://doi.org/10.1103/PhysRevE.104.054115} {\bibfield
  {journal} {\bibinfo  {journal} {Phys. Rev. E}\ }\textbf {\bibinfo {volume}
  {104}},\ \bibinfo {pages} {054115} (\bibinfo {year} {2021})}\BibitemShut
  {NoStop}%
\bibitem [{\citenamefont {Menczel}\ \emph {et~al.}(2021)\citenamefont
  {Menczel}, \citenamefont {Loisa}, \citenamefont {Brandner},\ and\
  \citenamefont {Flindt}}]{tur-satis}%
  \BibitemOpen
  \bibfield  {author} {\bibinfo {author} {\bibfnamefont {P.}~\bibnamefont
  {Menczel}}, \bibinfo {author} {\bibfnamefont {E.}~\bibnamefont {Loisa}},
  \bibinfo {author} {\bibfnamefont {K.}~\bibnamefont {Brandner}},\ and\
  \bibinfo {author} {\bibfnamefont {C.}~\bibnamefont {Flindt}},\ }\bibfield
  {title} {\bibinfo {title} {Thermodynamic uncertainty relations for coherently
  driven open quantum systems},\ }\href
  {https://doi.org/https://doi.org/10.48550/arXiv.2104.12712} {\bibfield
  {journal} {\bibinfo  {journal} {J. Phys. A: Math. Theor.}\ }\textbf {\bibinfo
  {volume} {54}},\ \bibinfo {pages} {314002} (\bibinfo {year}
  {2021})}\BibitemShut {NoStop}%
\bibitem [{\citenamefont {{Van Vu}}\ and\ \citenamefont
  {{Saito}}(2021)}]{coherent-5}%
  \BibitemOpen
  \bibfield  {author} {\bibinfo {author} {\bibfnamefont {T.}~\bibnamefont {{Van
  Vu}}}\ and\ \bibinfo {author} {\bibfnamefont {K.}~\bibnamefont {{Saito}}},\
  }\bibfield  {title} {\bibinfo {title} {{Thermodynamics of Precision in
  Markovian Open Quantum Dynamics}},\ }\href@noop {} {\bibfield  {journal}
  {\bibinfo  {journal} {arXiv e-prints}\ ,\ \bibinfo {eid} {arXiv:2111.04599}}
  (\bibinfo {year} {2021})},\ \Eprint {https://arxiv.org/abs/2111.04599}
  {arXiv:2111.04599 [cond-mat.stat-mech]} \BibitemShut {NoStop}%
\bibitem [{\citenamefont {Rignon-Bret}\ \emph {et~al.}(2021)\citenamefont
  {Rignon-Bret}, \citenamefont {Guarnieri}, \citenamefont {Goold},\ and\
  \citenamefont {Mitchison}}]{coherent-7}%
  \BibitemOpen
  \bibfield  {author} {\bibinfo {author} {\bibfnamefont {A.}~\bibnamefont
  {Rignon-Bret}}, \bibinfo {author} {\bibfnamefont {G.}~\bibnamefont
  {Guarnieri}}, \bibinfo {author} {\bibfnamefont {J.}~\bibnamefont {Goold}},\
  and\ \bibinfo {author} {\bibfnamefont {M.~T.}\ \bibnamefont {Mitchison}},\
  }\bibfield  {title} {\bibinfo {title} {Thermodynamics of precision in quantum
  nanomachines},\ }\href {https://doi.org/10.1103/PhysRevE.103.012133}
  {\bibfield  {journal} {\bibinfo  {journal} {Phys. Rev. E}\ }\textbf {\bibinfo
  {volume} {103}},\ \bibinfo {pages} {012133} (\bibinfo {year}
  {2021})}\BibitemShut {NoStop}%
\bibitem [{\citenamefont {Miller}\ \emph {et~al.}(2021)\citenamefont {Miller},
  \citenamefont {Mohammady}, \citenamefont {Perarnau-Llobet},\ and\
  \citenamefont {Guarnieri}}]{miller-slow}%
  \BibitemOpen
  \bibfield  {author} {\bibinfo {author} {\bibfnamefont {H.~J.~D.}\
  \bibnamefont {Miller}}, \bibinfo {author} {\bibfnamefont {M.~H.}\
  \bibnamefont {Mohammady}}, \bibinfo {author} {\bibfnamefont {M.}~\bibnamefont
  {Perarnau-Llobet}},\ and\ \bibinfo {author} {\bibfnamefont {G.}~\bibnamefont
  {Guarnieri}},\ }\bibfield  {title} {\bibinfo {title} {Thermodynamic
  uncertainty relation in slowly driven quantum heat engines},\ }\href
  {https://doi.org/10.1103/PhysRevLett.126.210603} {\bibfield  {journal}
  {\bibinfo  {journal} {Phys. Rev. Lett.}\ }\textbf {\bibinfo {volume} {126}},\
  \bibinfo {pages} {210603} (\bibinfo {year} {2021})}\BibitemShut {NoStop}%
\bibitem [{\citenamefont {Lee}\ \emph {et~al.}(2021)\citenamefont {Lee},
  \citenamefont {Ha},\ and\ \citenamefont {Jeong}}]{lee-otto-tur}%
  \BibitemOpen
  \bibfield  {author} {\bibinfo {author} {\bibfnamefont {S.}~\bibnamefont
  {Lee}}, \bibinfo {author} {\bibfnamefont {M.}~\bibnamefont {Ha}},\ and\
  \bibinfo {author} {\bibfnamefont {H.}~\bibnamefont {Jeong}},\ }\bibfield
  {title} {\bibinfo {title} {Quantumness and thermodynamic uncertainty relation
  of the finite-time otto cycle},\ }\href
  {https://doi.org/10.1103/PhysRevE.103.022136} {\bibfield  {journal} {\bibinfo
   {journal} {Phys. Rev. E}\ }\textbf {\bibinfo {volume} {103}},\ \bibinfo
  {pages} {022136} (\bibinfo {year} {2021})}\BibitemShut {NoStop}%
\bibitem [{\citenamefont {Sacchi}(2021)}]{sachhi-otto-tur}%
  \BibitemOpen
  \bibfield  {author} {\bibinfo {author} {\bibfnamefont {M.~F.}\ \bibnamefont
  {Sacchi}},\ }\bibfield  {title} {\bibinfo {title} {Thermodynamic uncertainty
  relations for bosonic otto engines},\ }\href
  {https://doi.org/10.1103/PhysRevE.103.012111} {\bibfield  {journal} {\bibinfo
   {journal} {Phys. Rev. E}\ }\textbf {\bibinfo {volume} {103}},\ \bibinfo
  {pages} {012111} (\bibinfo {year} {2021})}\BibitemShut {NoStop}%
\bibitem [{\citenamefont {Gelbwaser-Klimovsky}\ \emph
  {et~al.}(2013)\citenamefont {Gelbwaser-Klimovsky}, \citenamefont {Alicki},\
  and\ \citenamefont {Kurizki}}]{klimovsky13minimal}%
  \BibitemOpen
  \bibfield  {author} {\bibinfo {author} {\bibfnamefont {D.}~\bibnamefont
  {Gelbwaser-Klimovsky}}, \bibinfo {author} {\bibfnamefont {R.}~\bibnamefont
  {Alicki}},\ and\ \bibinfo {author} {\bibfnamefont {G.}~\bibnamefont
  {Kurizki}},\ }\bibfield  {title} {\bibinfo {title} {Minimal universal quantum
  heat machine},\ }\href {https://doi.org/10.1103/PhysRevE.87.012140}
  {\bibfield  {journal} {\bibinfo  {journal} {Phys. Rev. E}\ }\textbf {\bibinfo
  {volume} {87}},\ \bibinfo {pages} {012140} (\bibinfo {year}
  {2013})}\BibitemShut {NoStop}%
\bibitem [{\citenamefont {Shirley}(1965)}]{floquet-1}%
  \BibitemOpen
  \bibfield  {author} {\bibinfo {author} {\bibfnamefont {J.~H.}\ \bibnamefont
  {Shirley}},\ }\bibfield  {title} {\bibinfo {title} {Solution of the
  schr\"odinger equation with a hamiltonian periodic in time},\ }\href
  {https://doi.org/10.1103/PhysRev.138.B979} {\bibfield  {journal} {\bibinfo
  {journal} {Phys. Rev.}\ }\textbf {\bibinfo {volume} {138}},\ \bibinfo {pages}
  {B979} (\bibinfo {year} {1965})}\BibitemShut {NoStop}%
\bibitem [{\citenamefont {Sambe}(1973)}]{floquet-2}%
  \BibitemOpen
  \bibfield  {author} {\bibinfo {author} {\bibfnamefont {H.}~\bibnamefont
  {Sambe}},\ }\bibfield  {title} {\bibinfo {title} {Steady states and
  quasienergies of a quantum-mechanical system in an oscillating field},\
  }\href {https://doi.org/10.1103/PhysRevA.7.2203} {\bibfield  {journal}
  {\bibinfo  {journal} {Phys. Rev. A}\ }\textbf {\bibinfo {volume} {7}},\
  \bibinfo {pages} {2203} (\bibinfo {year} {1973})}\BibitemShut {NoStop}%
\bibitem [{\citenamefont {Kohler}\ \emph {et~al.}(1997)\citenamefont {Kohler},
  \citenamefont {Dittrich},\ and\ \citenamefont {H\"anggi}}]{floquet-3}%
  \BibitemOpen
  \bibfield  {author} {\bibinfo {author} {\bibfnamefont {S.}~\bibnamefont
  {Kohler}}, \bibinfo {author} {\bibfnamefont {T.}~\bibnamefont {Dittrich}},\
  and\ \bibinfo {author} {\bibfnamefont {P.}~\bibnamefont {H\"anggi}},\
  }\bibfield  {title} {\bibinfo {title} {Floquet-markovian description of the
  parametrically driven, dissipative harmonic quantum oscillator},\ }\href
  {https://doi.org/10.1103/PhysRevE.55.300} {\bibfield  {journal} {\bibinfo
  {journal} {Phys. Rev. E}\ }\textbf {\bibinfo {volume} {55}},\ \bibinfo
  {pages} {300} (\bibinfo {year} {1997})}\BibitemShut {NoStop}%
\bibitem [{\citenamefont {Grifoni}\ and\ \citenamefont
  {Hänggi}(1998)}]{floquet-4}%
  \BibitemOpen
  \bibfield  {author} {\bibinfo {author} {\bibfnamefont {M.}~\bibnamefont
  {Grifoni}}\ and\ \bibinfo {author} {\bibfnamefont {P.}~\bibnamefont
  {Hänggi}},\ }\bibfield  {title} {\bibinfo {title} {Driven quantum
  tunneling},\ }\href
  {https://doi.org/https://doi.org/10.1016/S0370-1573(98)00022-2} {\bibfield
  {journal} {\bibinfo  {journal} {Physics Reports}\ }\textbf {\bibinfo {volume}
  {304}},\ \bibinfo {pages} {229} (\bibinfo {year} {1998})}\BibitemShut
  {NoStop}%
\bibitem [{\citenamefont {Ito}\ \emph {et~al.}(2019)\citenamefont {Ito},
  \citenamefont {Jiang},\ and\ \citenamefont {Watanabe}}]{ito2019universal}%
  \BibitemOpen
  \bibfield  {author} {\bibinfo {author} {\bibfnamefont {K.}~\bibnamefont
  {Ito}}, \bibinfo {author} {\bibfnamefont {C.}~\bibnamefont {Jiang}},\ and\
  \bibinfo {author} {\bibfnamefont {G.}~\bibnamefont {Watanabe}},\ }\href@noop
  {} {\bibinfo {title} {Universal bounds for fluctuations in small heat
  engines}} (\bibinfo {year} {2019}),\ \Eprint
  {https://arxiv.org/abs/1910.08096} {arXiv:1910.08096 [cond-mat.stat-mech]}
  \BibitemShut {NoStop}%
\bibitem [{\citenamefont {Saryal}\ \emph
  {et~al.}(2021{\natexlab{b}})\citenamefont {Saryal}, \citenamefont {Gerry},
  \citenamefont {Khait}, \citenamefont {Segal},\ and\ \citenamefont
  {Agarwalla}}]{Bijay-machine-ss}%
  \BibitemOpen
  \bibfield  {author} {\bibinfo {author} {\bibfnamefont {S.}~\bibnamefont
  {Saryal}}, \bibinfo {author} {\bibfnamefont {M.}~\bibnamefont {Gerry}},
  \bibinfo {author} {\bibfnamefont {I.}~\bibnamefont {Khait}}, \bibinfo
  {author} {\bibfnamefont {D.}~\bibnamefont {Segal}},\ and\ \bibinfo {author}
  {\bibfnamefont {B.~K.}\ \bibnamefont {Agarwalla}},\ }\bibfield  {title}
  {\bibinfo {title} {Universal bounds on fluctuations in continuous thermal
  machines},\ }\href {https://doi.org/10.1103/PhysRevLett.127.190603}
  {\bibfield  {journal} {\bibinfo  {journal} {Phys. Rev. Lett.}\ }\textbf
  {\bibinfo {volume} {127}},\ \bibinfo {pages} {190603} (\bibinfo {year}
  {2021}{\natexlab{b}})}\BibitemShut {NoStop}%
\bibitem [{\citenamefont {Saryal}\ and\ \citenamefont
  {Agarwalla}(2021)}]{Bijay-Otto}%
  \BibitemOpen
  \bibfield  {author} {\bibinfo {author} {\bibfnamefont {S.}~\bibnamefont
  {Saryal}}\ and\ \bibinfo {author} {\bibfnamefont {B.~K.}\ \bibnamefont
  {Agarwalla}},\ }\bibfield  {title} {\bibinfo {title} {Bounds on fluctuations
  for finite-time quantum otto cycle},\ }\href
  {https://doi.org/10.1103/PhysRevE.103.L060103} {\bibfield  {journal}
  {\bibinfo  {journal} {Phys. Rev. E}\ }\textbf {\bibinfo {volume} {103}},\
  \bibinfo {pages} {L060103} (\bibinfo {year} {2021})}\BibitemShut {NoStop}%
\bibitem [{\citenamefont {Gasparinetti}\ \emph {et~al.}(2013)\citenamefont
  {Gasparinetti}, \citenamefont {Solinas}, \citenamefont {Pugnetti},
  \citenamefont {Fazio},\ and\ \citenamefont {Pekola}}]{secular-1}%
  \BibitemOpen
  \bibfield  {author} {\bibinfo {author} {\bibfnamefont {S.}~\bibnamefont
  {Gasparinetti}}, \bibinfo {author} {\bibfnamefont {P.}~\bibnamefont
  {Solinas}}, \bibinfo {author} {\bibfnamefont {S.}~\bibnamefont {Pugnetti}},
  \bibinfo {author} {\bibfnamefont {R.}~\bibnamefont {Fazio}},\ and\ \bibinfo
  {author} {\bibfnamefont {J.~P.}\ \bibnamefont {Pekola}},\ }\bibfield  {title}
  {\bibinfo {title} {Environment-governed dynamics in driven quantum systems},\
  }\href {https://doi.org/https://doi.org/10.1103/PhysRevLett.110.150403}
  {\bibfield  {journal} {\bibinfo  {journal} {Phys. Rev. Lett.}\ }\textbf
  {\bibinfo {volume} {110}},\ \bibinfo {pages} {150403} (\bibinfo {year}
  {2013})}\BibitemShut {NoStop}%
\bibitem [{\citenamefont {Gasparinetti}\ \emph {et~al.}(2014)\citenamefont
  {Gasparinetti}, \citenamefont {Solinas}, \citenamefont {Braggio},\ and\
  \citenamefont {Sassetti}}]{secular-2}%
  \BibitemOpen
  \bibfield  {author} {\bibinfo {author} {\bibfnamefont {S.}~\bibnamefont
  {Gasparinetti}}, \bibinfo {author} {\bibfnamefont {P.}~\bibnamefont
  {Solinas}}, \bibinfo {author} {\bibfnamefont {A.}~\bibnamefont {Braggio}},\
  and\ \bibinfo {author} {\bibfnamefont {M.}~\bibnamefont {Sassetti}},\
  }\bibfield  {title} {\bibinfo {title} {Heat-exchange statistics in driven
  open quantum systems},\ }\href
  {https://doi.org/https://doi.org/10.1088/1367-2630/16/11/115001} {\bibfield
  {journal} {\bibinfo  {journal} {New J. Phys.}\ }\textbf {\bibinfo {volume}
  {16}},\ \bibinfo {pages} {115001} (\bibinfo {year} {2014})}\BibitemShut
  {NoStop}%
\bibitem [{\citenamefont {Breuer}\ and\ \citenamefont
  {Petruccione}(2002)}]{breuer02}%
  \BibitemOpen
  \bibfield  {author} {\bibinfo {author} {\bibfnamefont {H.~P.}\ \bibnamefont
  {Breuer}}\ and\ \bibinfo {author} {\bibfnamefont {F.}~\bibnamefont
  {Petruccione}},\ }\href@noop {} {\emph {\bibinfo {title} {The Theory of Open
  Quantum Systems}}}\ (\bibinfo  {publisher} {Oxford University Press},\
  \bibinfo {year} {2002})\BibitemShut {NoStop}%
\bibitem [{\citenamefont {Alicki}(2014)}]{alicki14quantum}%
  \BibitemOpen
  \bibfield  {author} {\bibinfo {author} {\bibfnamefont {R.}~\bibnamefont
  {Alicki}},\ }\bibfield  {title} {\bibinfo {title} {Quantum thermodynamics. an
  example of two-level quantum machine},\ }\href
  {https://doi.org/https://doi.org/10.1142/S1230161214400022} {\bibfield
  {journal} {\bibinfo  {journal} {Open Systems And Information Dynamics}\
  }\textbf {\bibinfo {volume} {21}},\ \bibinfo {pages} {1440002} (\bibinfo
  {year} {2014})}\BibitemShut {NoStop}%
\bibitem [{\citenamefont {Shahmoon}\ and\ \citenamefont
  {Kurizki}(2013)}]{squeez-1}%
  \BibitemOpen
  \bibfield  {author} {\bibinfo {author} {\bibfnamefont {E.}~\bibnamefont
  {Shahmoon}}\ and\ \bibinfo {author} {\bibfnamefont {G.}~\bibnamefont
  {Kurizki}},\ }\bibfield  {title} {\bibinfo {title} {Engineering a thermal
  squeezed reservoir by energy-level modulation},\ }\href
  {https://doi.org/10.1103/PhysRevA.87.013841} {\bibfield  {journal} {\bibinfo
  {journal} {Phys. Rev. A}\ }\textbf {\bibinfo {volume} {87}},\ \bibinfo
  {pages} {013841} (\bibinfo {year} {2013})}\BibitemShut {NoStop}%
\bibitem [{\citenamefont {Niedenzu}\ \emph {et~al.}(2019)\citenamefont
  {Niedenzu}, \citenamefont {Huber},\ and\ \citenamefont
  {Boukobza}}]{niedenzu19concepts}%
  \BibitemOpen
  \bibfield  {author} {\bibinfo {author} {\bibfnamefont {W.}~\bibnamefont
  {Niedenzu}}, \bibinfo {author} {\bibfnamefont {M.}~\bibnamefont {Huber}},\
  and\ \bibinfo {author} {\bibfnamefont {E.}~\bibnamefont {Boukobza}},\
  }\bibfield  {title} {\bibinfo {title} {Concepts of work in autonomous quantum
  heat engines},\ }\href {https://doi.org/10.22331/q-2019-10-14-195} {\bibfield
   {journal} {\bibinfo  {journal} {{Quantum}}\ }\textbf {\bibinfo {volume}
  {3}},\ \bibinfo {pages} {195} (\bibinfo {year} {2019})}\BibitemShut {NoStop}%
\bibitem [{\citenamefont {Schaller}(2014)}]{schaller2014open}%
  \BibitemOpen
  \bibfield  {author} {\bibinfo {author} {\bibfnamefont {G.}~\bibnamefont
  {Schaller}},\ }\href {https://books.google.pl/books?id=V8IJngEACAAJ} {\emph
  {\bibinfo {title} {Open Quantum Systems Far from Equilibrium}}},\ Lecture
  Notes in Physics\ (\bibinfo  {publisher} {Springer International
  Publishing},\ \bibinfo {year} {2014})\BibitemShut {NoStop}%
\bibitem [{\citenamefont {Alicki}\ and\ \citenamefont
  {Lendi}(2007)}]{alicki2007quantum}%
  \BibitemOpen
  \bibfield  {author} {\bibinfo {author} {\bibfnamefont {R.}~\bibnamefont
  {Alicki}}\ and\ \bibinfo {author} {\bibfnamefont {K.}~\bibnamefont {Lendi}},\
  }\href {https://books.google.pl/books?id=Y9NrCQAAQBAJ} {\emph {\bibinfo
  {title} {Quantum Dynamical Semigroups and Applications}}},\ Lecture Notes in
  Physics\ (\bibinfo  {publisher} {Springer Berlin Heidelberg},\ \bibinfo
  {year} {2007})\BibitemShut {NoStop}%
\bibitem [{\citenamefont {Gelbwaser-Klimovsky}\ \emph
  {et~al.}(2015)\citenamefont {Gelbwaser-Klimovsky}, \citenamefont {Niedenzu},\
  and\ \citenamefont {Kurizki}}]{klimovsky15thermodynamics}%
  \BibitemOpen
  \bibfield  {author} {\bibinfo {author} {\bibfnamefont {D.}~\bibnamefont
  {Gelbwaser-Klimovsky}}, \bibinfo {author} {\bibfnamefont {W.}~\bibnamefont
  {Niedenzu}},\ and\ \bibinfo {author} {\bibfnamefont {G.}~\bibnamefont
  {Kurizki}},\ }\bibfield  {title} {\bibinfo {title} {Chapter twelve -
  thermodynamics of quantum systems under dynamical control},\ }\href
  {https://doi.org/http://dx.doi.org/10.1016/bs.aamop.2015.07.002} {\bibfield
  {journal} {\bibinfo  {journal} {Advances In Atomic, Molecular, and Optical
  Physics}\ }\textbf {\bibinfo {volume} {64}},\ \bibinfo {pages} {329 }
  (\bibinfo {year} {2015})}\BibitemShut {NoStop}%
\bibitem [{\citenamefont {Kosloff}\ and\ \citenamefont
  {Rezek}(2017)}]{kosloff17the}%
  \BibitemOpen
  \bibfield  {author} {\bibinfo {author} {\bibfnamefont {R.}~\bibnamefont
  {Kosloff}}\ and\ \bibinfo {author} {\bibfnamefont {Y.}~\bibnamefont
  {Rezek}},\ }\bibfield  {title} {\bibinfo {title} {The quantum harmonic otto
  cycle},\ }\href {https://doi.org/https://doi.org/10.3390/e19040136}
  {\bibfield  {journal} {\bibinfo  {journal} {Entropy}\ }\textbf {\bibinfo
  {volume} {19}},\ \bibinfo {pages} {136} (\bibinfo {year} {2017})}\BibitemShut
  {NoStop}%
\bibitem [{\citenamefont {Seifert}(2018)}]{tur-satis2}%
  \BibitemOpen
  \bibfield  {author} {\bibinfo {author} {\bibfnamefont {U.}~\bibnamefont
  {Seifert}},\ }\bibfield  {title} {\bibinfo {title} {Stochastic
  thermodynamics: From principles to the cost of precision},\ }\href
  {https://doi.org/https://doi.org/10.1016/j.physa.2017.10.024} {\bibfield
  {journal} {\bibinfo  {journal} {Physica A}\ }\textbf {\bibinfo {volume}
  {504}},\ \bibinfo {pages} {176} (\bibinfo {year} {2018})}\BibitemShut
  {NoStop}%
\bibitem [{\citenamefont {Caneva}\ \emph {et~al.}(2011)\citenamefont {Caneva},
  \citenamefont {Calarco},\ and\ \citenamefont
  {Montangero}}]{caneva2011chopped}%
  \BibitemOpen
  \bibfield  {author} {\bibinfo {author} {\bibfnamefont {T.}~\bibnamefont
  {Caneva}}, \bibinfo {author} {\bibfnamefont {T.}~\bibnamefont {Calarco}},\
  and\ \bibinfo {author} {\bibfnamefont {S.}~\bibnamefont {Montangero}},\
  }\bibfield  {title} {\bibinfo {title} {Chopped random-basis quantum
  optimization},\ }\href@noop {} {\bibfield  {journal} {\bibinfo  {journal}
  {Physical Review A}\ }\textbf {\bibinfo {volume} {84}},\ \bibinfo {pages}
  {022326} (\bibinfo {year} {2011})}\BibitemShut {NoStop}%
\bibitem [{\citenamefont {Doria}\ \emph {et~al.}(2011)\citenamefont {Doria},
  \citenamefont {Calarco},\ and\ \citenamefont
  {Montangero}}]{doria2011optimal}%
  \BibitemOpen
  \bibfield  {author} {\bibinfo {author} {\bibfnamefont {P.}~\bibnamefont
  {Doria}}, \bibinfo {author} {\bibfnamefont {T.}~\bibnamefont {Calarco}},\
  and\ \bibinfo {author} {\bibfnamefont {S.}~\bibnamefont {Montangero}},\
  }\bibfield  {title} {\bibinfo {title} {Optimal control technique for
  many-body quantum dynamics},\ }\href@noop {} {\bibfield  {journal} {\bibinfo
  {journal} {Physical review letters}\ }\textbf {\bibinfo {volume} {106}},\
  \bibinfo {pages} {190501} (\bibinfo {year} {2011})}\BibitemShut {NoStop}%
\bibitem [{\citenamefont {Müller}\ \emph {et~al.}(2022)\citenamefont
  {Müller}, \citenamefont {Said}, \citenamefont {Jelezko}, \citenamefont
  {Calarco},\ and\ \citenamefont {Montangero}}]{muller_2022}%
  \BibitemOpen
  \bibfield  {author} {\bibinfo {author} {\bibfnamefont {M.~M.}\ \bibnamefont
  {Müller}}, \bibinfo {author} {\bibfnamefont {R.~S.}\ \bibnamefont {Said}},
  \bibinfo {author} {\bibfnamefont {F.}~\bibnamefont {Jelezko}}, \bibinfo
  {author} {\bibfnamefont {T.}~\bibnamefont {Calarco}},\ and\ \bibinfo {author}
  {\bibfnamefont {S.}~\bibnamefont {Montangero}},\ }\bibfield  {title}
  {\bibinfo {title} {One decade of quantum optimal control in the chopped
  random basis},\ }\href {https://doi.org/10.1088/1361-6633/ac723c} {\bibfield
  {journal} {\bibinfo  {journal} {Reports on Progress in Physics}\ }\textbf
  {\bibinfo {volume} {85}},\ \bibinfo {pages} {076001} (\bibinfo {year}
  {2022})}\BibitemShut {NoStop}%
\bibitem [{\citenamefont {Mukherjee}\ \emph {et~al.}(2013)\citenamefont
  {Mukherjee}, \citenamefont {Carlini}, \citenamefont {Mari}, \citenamefont
  {Caneva}, \citenamefont {Montangero}, \citenamefont {Calarco}, \citenamefont
  {Fazio},\ and\ \citenamefont {Giovannetti}}]{mukherjee13speeding}%
  \BibitemOpen
  \bibfield  {author} {\bibinfo {author} {\bibfnamefont {V.}~\bibnamefont
  {Mukherjee}}, \bibinfo {author} {\bibfnamefont {A.}~\bibnamefont {Carlini}},
  \bibinfo {author} {\bibfnamefont {A.}~\bibnamefont {Mari}}, \bibinfo {author}
  {\bibfnamefont {T.}~\bibnamefont {Caneva}}, \bibinfo {author} {\bibfnamefont
  {S.}~\bibnamefont {Montangero}}, \bibinfo {author} {\bibfnamefont
  {T.}~\bibnamefont {Calarco}}, \bibinfo {author} {\bibfnamefont
  {R.}~\bibnamefont {Fazio}},\ and\ \bibinfo {author} {\bibfnamefont
  {V.}~\bibnamefont {Giovannetti}},\ }\bibfield  {title} {\bibinfo {title}
  {Speeding up and slowing down the relaxation of a qubit by optimal control},\
  }\href {https://doi.org/10.1103/PhysRevA.88.062326} {\bibfield  {journal}
  {\bibinfo  {journal} {Phys. Rev. A}\ }\textbf {\bibinfo {volume} {88}},\
  \bibinfo {pages} {062326} (\bibinfo {year} {2013})}\BibitemShut {NoStop}%
\bibitem [{\citenamefont {Caneva}\ \emph {et~al.}(2014)\citenamefont {Caneva},
  \citenamefont {Silva}, \citenamefont {Fazio}, \citenamefont {Lloyd},
  \citenamefont {Calarco},\ and\ \citenamefont
  {Montangero}}]{caneva14complexity}%
  \BibitemOpen
  \bibfield  {author} {\bibinfo {author} {\bibfnamefont {T.}~\bibnamefont
  {Caneva}}, \bibinfo {author} {\bibfnamefont {A.}~\bibnamefont {Silva}},
  \bibinfo {author} {\bibfnamefont {R.}~\bibnamefont {Fazio}}, \bibinfo
  {author} {\bibfnamefont {S.}~\bibnamefont {Lloyd}}, \bibinfo {author}
  {\bibfnamefont {T.}~\bibnamefont {Calarco}},\ and\ \bibinfo {author}
  {\bibfnamefont {S.}~\bibnamefont {Montangero}},\ }\bibfield  {title}
  {\bibinfo {title} {Complexity of controlling quantum many-body dynamics},\
  }\href {https://doi.org/10.1103/PhysRevA.89.042322} {\bibfield  {journal}
  {\bibinfo  {journal} {Phys. Rev. A}\ }\textbf {\bibinfo {volume} {89}},\
  \bibinfo {pages} {042322} (\bibinfo {year} {2014})}\BibitemShut {NoStop}%
\bibitem [{\citenamefont {Omran}\ \emph {et~al.}(2019)\citenamefont {Omran},
  \citenamefont {Levine}, \citenamefont {Keesling}, \citenamefont {Semeghini},
  \citenamefont {Wang}, \citenamefont {Ebadi}, \citenamefont {Bernien},
  \citenamefont {Zibrov}, \citenamefont {Pichler}, \citenamefont {Choi},
  \citenamefont {Cui}, \citenamefont {Rossignolo}, \citenamefont {Rembold},
  \citenamefont {Montangero}, \citenamefont {Calarco}, \citenamefont {Endres},
  \citenamefont {Greiner}, \citenamefont {Vuletić},\ and\ \citenamefont
  {Lukin}}]{omran19generation}%
  \BibitemOpen
  \bibfield  {author} {\bibinfo {author} {\bibfnamefont {A.}~\bibnamefont
  {Omran}}, \bibinfo {author} {\bibfnamefont {H.}~\bibnamefont {Levine}},
  \bibinfo {author} {\bibfnamefont {A.}~\bibnamefont {Keesling}}, \bibinfo
  {author} {\bibfnamefont {G.}~\bibnamefont {Semeghini}}, \bibinfo {author}
  {\bibfnamefont {T.~T.}\ \bibnamefont {Wang}}, \bibinfo {author}
  {\bibfnamefont {S.}~\bibnamefont {Ebadi}}, \bibinfo {author} {\bibfnamefont
  {H.}~\bibnamefont {Bernien}}, \bibinfo {author} {\bibfnamefont {A.~S.}\
  \bibnamefont {Zibrov}}, \bibinfo {author} {\bibfnamefont {H.}~\bibnamefont
  {Pichler}}, \bibinfo {author} {\bibfnamefont {S.}~\bibnamefont {Choi}},
  \bibinfo {author} {\bibfnamefont {J.}~\bibnamefont {Cui}}, \bibinfo {author}
  {\bibfnamefont {M.}~\bibnamefont {Rossignolo}}, \bibinfo {author}
  {\bibfnamefont {P.}~\bibnamefont {Rembold}}, \bibinfo {author} {\bibfnamefont
  {S.}~\bibnamefont {Montangero}}, \bibinfo {author} {\bibfnamefont
  {T.}~\bibnamefont {Calarco}}, \bibinfo {author} {\bibfnamefont
  {M.}~\bibnamefont {Endres}}, \bibinfo {author} {\bibfnamefont
  {M.}~\bibnamefont {Greiner}}, \bibinfo {author} {\bibfnamefont
  {V.}~\bibnamefont {Vuletić}},\ and\ \bibinfo {author} {\bibfnamefont
  {M.~D.}\ \bibnamefont {Lukin}},\ }\bibfield  {title} {\bibinfo {title}
  {Generation and manipulation of {S}chr{\"{o}}dinger cat states in {R}ydberg
  atom arrays},\ }\href {https://doi.org/10.1126/science.aax9743} {\bibfield
  {journal} {\bibinfo  {journal} {Science}\ }\textbf {\bibinfo {volume}
  {365}},\ \bibinfo {pages} {570} (\bibinfo {year} {2019})}\BibitemShut
  {NoStop}%
\bibitem [{\citenamefont {Borselli}\ \emph {et~al.}(2021)\citenamefont
  {Borselli}, \citenamefont {Maiw\"oger}, \citenamefont {Zhang}, \citenamefont
  {Haslinger}, \citenamefont {Mukherjee}, \citenamefont {Negretti},
  \citenamefont {Montangero}, \citenamefont {Calarco}, \citenamefont {Mazets},
  \citenamefont {Bonneau},\ and\ \citenamefont {Schmiedmayer}}]{borselli21two}%
  \BibitemOpen
  \bibfield  {author} {\bibinfo {author} {\bibfnamefont {F.}~\bibnamefont
  {Borselli}}, \bibinfo {author} {\bibfnamefont {M.}~\bibnamefont
  {Maiw\"oger}}, \bibinfo {author} {\bibfnamefont {T.}~\bibnamefont {Zhang}},
  \bibinfo {author} {\bibfnamefont {P.}~\bibnamefont {Haslinger}}, \bibinfo
  {author} {\bibfnamefont {V.}~\bibnamefont {Mukherjee}}, \bibinfo {author}
  {\bibfnamefont {A.}~\bibnamefont {Negretti}}, \bibinfo {author}
  {\bibfnamefont {S.}~\bibnamefont {Montangero}}, \bibinfo {author}
  {\bibfnamefont {T.}~\bibnamefont {Calarco}}, \bibinfo {author} {\bibfnamefont
  {I.}~\bibnamefont {Mazets}}, \bibinfo {author} {\bibfnamefont
  {M.}~\bibnamefont {Bonneau}},\ and\ \bibinfo {author} {\bibfnamefont
  {J.}~\bibnamefont {Schmiedmayer}},\ }\bibfield  {title} {\bibinfo {title}
  {Two-particle interference with double twin-atom beams},\ }\href
  {https://doi.org/10.1103/PhysRevLett.126.083603} {\bibfield  {journal}
  {\bibinfo  {journal} {Phys. Rev. Lett.}\ }\textbf {\bibinfo {volume} {126}},\
  \bibinfo {pages} {083603} (\bibinfo {year} {2021})}\BibitemShut {NoStop}%
\bibitem [{\citenamefont {Kosloff}\ and\ \citenamefont
  {Feldmann}(2002)}]{kosloff02discrete}%
  \BibitemOpen
  \bibfield  {author} {\bibinfo {author} {\bibfnamefont {R.}~\bibnamefont
  {Kosloff}}\ and\ \bibinfo {author} {\bibfnamefont {T.}~\bibnamefont
  {Feldmann}},\ }\bibfield  {title} {\bibinfo {title} {Discrete four-stroke
  quantum heat engine exploring the origin of friction},\ }\href
  {https://doi.org/10.1103/PhysRevE.65.055102} {\bibfield  {journal} {\bibinfo
  {journal} {Phys. Rev. E}\ }\textbf {\bibinfo {volume} {65}},\ \bibinfo
  {pages} {055102} (\bibinfo {year} {2002})}\BibitemShut {NoStop}%
\bibitem [{\citenamefont {Mukherjee}\ \emph {et~al.}(2016)\citenamefont
  {Mukherjee}, \citenamefont {Niedenzu}, \citenamefont {Kofman},\ and\
  \citenamefont {Kurizki}}]{mukherjee16speed}%
  \BibitemOpen
  \bibfield  {author} {\bibinfo {author} {\bibfnamefont {V.}~\bibnamefont
  {Mukherjee}}, \bibinfo {author} {\bibfnamefont {W.}~\bibnamefont {Niedenzu}},
  \bibinfo {author} {\bibfnamefont {A.~G.}\ \bibnamefont {Kofman}},\ and\
  \bibinfo {author} {\bibfnamefont {G.}~\bibnamefont {Kurizki}},\ }\bibfield
  {title} {\bibinfo {title} {Speed and efficiency limits of multilevel
  incoherent heat engines},\ }\href
  {https://doi.org/10.1103/PhysRevE.94.062109} {\bibfield  {journal} {\bibinfo
  {journal} {Phys. Rev. E}\ }\textbf {\bibinfo {volume} {94}},\ \bibinfo
  {pages} {062109} (\bibinfo {year} {2016})}\BibitemShut {NoStop}%
\bibitem [{\citenamefont {Szczygielski}\ \emph {et~al.}(2013)\citenamefont
  {Szczygielski}, \citenamefont {Gelbwaser-Klimovsky},\ and\ \citenamefont
  {Alicki}}]{szczygielski13}%
  \BibitemOpen
  \bibfield  {author} {\bibinfo {author} {\bibfnamefont {K.}~\bibnamefont
  {Szczygielski}}, \bibinfo {author} {\bibfnamefont {D.}~\bibnamefont
  {Gelbwaser-Klimovsky}},\ and\ \bibinfo {author} {\bibfnamefont
  {R.}~\bibnamefont {Alicki}},\ }\bibfield  {title} {\bibinfo {title}
  {Markovian master equation and thermodynamics of a two-level system in a
  strong laser field},\ }\href {https://doi.org/10.1103/PhysRevE.87.012120}
  {\bibfield  {journal} {\bibinfo  {journal} {Phys. Rev. E}\ }\textbf {\bibinfo
  {volume} {87}},\ \bibinfo {pages} {012120} (\bibinfo {year}
  {2013})}\BibitemShut {NoStop}%
\bibitem [{\citenamefont {Restrepo}\ \emph {et~al.}(2018)\citenamefont
  {Restrepo}, \citenamefont {Cerrillo}, \citenamefont {Strasberg},\ and\
  \citenamefont {Schaller}}]{secular-3}%
  \BibitemOpen
  \bibfield  {author} {\bibinfo {author} {\bibfnamefont {S.}~\bibnamefont
  {Restrepo}}, \bibinfo {author} {\bibfnamefont {J.}~\bibnamefont {Cerrillo}},
  \bibinfo {author} {\bibfnamefont {P.}~\bibnamefont {Strasberg}},\ and\
  \bibinfo {author} {\bibfnamefont {G.}~\bibnamefont {Schaller}},\ }\bibfield
  {title} {\bibinfo {title} {From quantum heat engines to laser cooling:
  Floquet theory beyond the born–markov approximation},\ }\href
  {https://doi.org/https://doi.org/10.1088/1367-2630/aac583} {\bibfield
  {journal} {\bibinfo  {journal} {New J. Phys.}\ }\textbf {\bibinfo {volume}
  {20}},\ \bibinfo {pages} {053063} (\bibinfo {year} {2018})}\BibitemShut
  {NoStop}%
\bibitem [{\citenamefont {Das}\ and\ \citenamefont {Mukherjee}(2020)}]{arpan}%
  \BibitemOpen
  \bibfield  {author} {\bibinfo {author} {\bibfnamefont {A.}~\bibnamefont
  {Das}}\ and\ \bibinfo {author} {\bibfnamefont {V.}~\bibnamefont
  {Mukherjee}},\ }\bibfield  {title} {\bibinfo {title} {Quantum-enhanced
  finite-time {O}tto cycle},\ }\href
  {https://doi.org/https://doi.org/10.1103/PhysRevResearch.2.033083} {\bibfield
   {journal} {\bibinfo  {journal} {Phys. Rev. Research}\ }\textbf {\bibinfo
  {volume} {2}},\ \bibinfo {pages} {033083} (\bibinfo {year}
  {2020})}\BibitemShut {NoStop}%
\bibitem [{\citenamefont {Mukherjee}\ \emph {et~al.}(2019)\citenamefont
  {Mukherjee}, \citenamefont {Zwick}, \citenamefont {Ghosh}, \citenamefont
  {Chen},\ and\ \citenamefont {Kurizki}}]{mukherjee19enhanced}%
  \BibitemOpen
  \bibfield  {author} {\bibinfo {author} {\bibfnamefont {V.}~\bibnamefont
  {Mukherjee}}, \bibinfo {author} {\bibfnamefont {A.}~\bibnamefont {Zwick}},
  \bibinfo {author} {\bibfnamefont {A.}~\bibnamefont {Ghosh}}, \bibinfo
  {author} {\bibfnamefont {X.}~\bibnamefont {Chen}},\ and\ \bibinfo {author}
  {\bibfnamefont {G.}~\bibnamefont {Kurizki}},\ }\bibfield  {title} {\bibinfo
  {title} {Enhanced precision bound of low-temperature quantum thermometry via
  dynamical control},\ }\href {https://doi.org/10.1038/s42005-019-0265-y}
  {\bibfield  {journal} {\bibinfo  {journal} {Commun Phys}\ }\textbf {\bibinfo
  {volume} {2}},\ \bibinfo {pages} {162} (\bibinfo {year} {2019})}\BibitemShut
  {NoStop}%
\end{thebibliography}%
\end{document}